%% file: ms.tex
\documentclass[12pt,preprint]{aastex}
\usepackage{float}

\shorttitle{X-ray Lines}
\shortauthors{Butler et al.}

\def\gtrsim{\mathrel{\hbox{\rlap{\hbox{\lower4pt\hbox{$\sim$}}}\hbox{$>$}}}}
\def\lessim{\mathrel{\hbox{\rlap{\hbox{\lower4pt\hbox{$\sim$}}}\hbox{$<$}}}}

\slugcomment{Submitted to ApJ}

\begin{document}

\title{On the Early Time X-ray Spectra of Swift Afterglows I: \\ Evidence for Anomalous Soft X-ray Emission}
\author{N.~R. Butler\altaffilmark{1,2}}
\altaffiltext{1}{Townes Fellow, Space Sciences Laboratory,
University of California, Berkeley, CA, 94720-7450, USA}
\altaffiltext{2}{Astronomy Department, University of California,
445 Campbell Hall, Berkeley, CA 94720-3411, USA}

\begin{abstract}

We have conducted a thorough and blind search for emission lines in 
$>70$ Swift X-ray afterglows of total exposure $\sim 10^7$s.
We find that most afterglows are consistent with pure power-laws plus
extinction.  Significant
outliers to the population exist at the 5-10\% level and have anomalously soft, possibly thermal spectra.  Four bursts
are singled out via possible detections of 2-5 lines: GRBs 060218, 060202, 050822, and 050714B.  
Alternatively, a blackbody model with $kT\sim 0.1-0.5$ keV can describe the soft emission in each afterglow.
The most significant
soft component detections in the full data set of $\sim 2000$ spectra correspond to GRB~060218/SN~2006aj, with line significances ranging
up to $\sim 20\sigma$.  A thermal plasma model fit to the data indicates that the flux is primarily due to L-shell transitions of
Fe at $\sim$ solar abundance.  
We associate ($>4\sigma$ significant) line triggers in
the 3 other events with K-shell transitions in light metals.  
We favor a model where the possible line emission in these afterglows arises from the mildly relativistic cocoon of matter surrounding the GRB
jet as it penetrates and exits the surface of the progenitor star.  
The emitting material in each burst is at a similar
distance $\sim 10^{12}-10^{13}$ cm, a similar density $\sim 10^{17}$ cm$^{-3}$, and subject to a similar flux of ionizing radiation.
The lines may correlate with the X-ray flaring.
For the blackbody interpretation, the soft flux may arise from break out of the GRB shock or plasma cocoon from the progenitor stellar wind, as
recently suggested for GRB~060218 \citep{campana06}.
Due to the low $z$ of GRB~060218, bursts faint in $\gamma-$rays with fluxes dominated by this soft X-ray component
could outnumber classical GRBs 100-1.  
\end{abstract}


\maketitle

\section{Introduction}
\label{sec:intro}

One of the key open questions in the study of Gamma-ray bursts (GRBs)
is that of the X-ray afterglow lines.  Claims of low to moderate 
significance emission lines have been made based on data from several
missions: Fe lines have been detected in afterglow data from {\it ASCA}
\citep{yoshida99}, {\it Beppo-SAX} \citep{piro99,antonelli00}, and
{\it Chandra} \citep{piro00}; lines from highly ionized light, 
multiple-$\alpha$ elements like
Mg, Si, S, Ar, and Ca have been detected in afterglow data from {\it XMM}
\citep{reeves02,watson03} and {\it Chandra} \citep{butler03}.  The
detections are challenging to explain because they typically imply large, 
concentrated masses of metals in the circumburst material 
\citep[see, e.g.,][]{lcg99} and a very efficient reprocessing of the 
non-thermal afterglow continuum into line radiation 
\citep[see,][]{bnr01,lrr02}.  

\citet{rutledge03} and \citet{skr05} argue that the claims made
to date lack the necessary significance needed to prove that the X-ray 
lines are real.  We address the line significance in the case of 
GRB~011211 in a recent paper \citep{butler05a} and find that the differing
significance estimates are simply due to different input assumption in the
continuum modeling.  To prove consistency between the analysis techniques
utilized by the \citet{reeves02} and \citet{rutledge03},
we developed software to autonomously detect one or multiple
emission lines.  Below, we employ the line search tools to comb the
vast {\it Swift X-ray Telescope}~(XRT) data set.  With nearly 0.4 years in total
exposure for $\gtrsim 70$ bursts observed typically within 1 hour of the GRB, the mounting
XRT sample represents a unique laboratory in which we can test the veracity of historical
line claims and probe the physics of new phenomena (e.g., flares, rapid and unusually flat light curves)
at early times.

As we discuss below, divided up in time, the majority ($\sim 90$\%) of {\it Swift}~X-ray afterglows are well
modeled by simple power-laws in energy, partially absorbed below $\sim 1$ keV
by gas along the line of sight.  The typical power-law photon index is $\Gamma \sim 2$.
However, a small fraction ($\sim 10$\%) of the afterglows in the
sample exhibit very different spectra ($\Gamma\sim 5$ power-laws or prominent residuals near 1 keV),
which trigger the line detection robot.  These spectra can be fit with a blackbody model
in addition to a power-law, or with emission lines in addition to the power-law.  One such event---
GRB~060218 \citep{cusumano06a}---is clearly associated with a SN~2006aj \citep{mirabal06,soller06,modjaz06}
and may be the smoking gun tying the anomalous soft X-ray spectra from this and 3 other GRB afterglows
to supernovae and possibly directly to the breakout of the GRB shock from the progenitor star.

\section{Data Reduction and Continuum Fits}

Our automated pipeline at Berkeley downloads the XRT data in near real
time.  The burst right ascension and
declination are gleaned from the Gamma-ray Coordinates Network (GCN)
notices in order to run the {\tt xrtpipeline} 
reduction task from the
HEAsoft~6.0\footnote{http://heasarc.gsfc.nasa.gov/docs/software/lheasoft/}
software release.  We use version 007 of the response matrices.
From there, we bin the data in time, exclude pileup
chip regions for each time interval, refine the X-ray position, and 
produce spectra using custom IDL scripts.  For the XRT position
counting (PC) mode pileup rejection, we discard regions with count rates 
in excess of 0.5 cts/s (0.5 - 10 keV) \citep[e.g.,][]{nousek06}.  For the Windowed 
Timing (WT) mode data the cutoff is 150 cts/s.  For each time slice,
we determined whether or not this threshold has been surpassed by
analyzing counts within a 3.5 pixel (inner radius) to 16 pixel (outer
radius) annulus around the source.  The inner radius is then expanded or
contracted to bring the count rate close to (but not beyond) the pileup 
threshold.  For $\sim 99$\% of the data by time (or $\sim 60$\% by mass), 
there is no pileup and the resulting inner radius is zero.

Spectral response files are generated using the {\tt xrtmkarf} task
for each time slice, and the time slices are weighted by exposure and
summed.
The spectra are fit in ISIS\footnote{http://space.mit.edu/CXC/ISIS/}.
For each spectral bin, we require a S/N of 3.5. 
We define S/N as the background-subtracted number of
counts divided by the square root of the sum of the signal counts and the
 variance in the background.  We define the background region as that 
 where the number of counts in an aperture the size of the source extraction
 region is within 2-sigma of the median background over the chip
 in that aperture for one contiguous follow-up observation.  In most
 cases, the background is entirely negligible ($\lessim 1$\% of the
 source flux).  

\begin{figure}[H]
\includegraphics[width=3.3in]{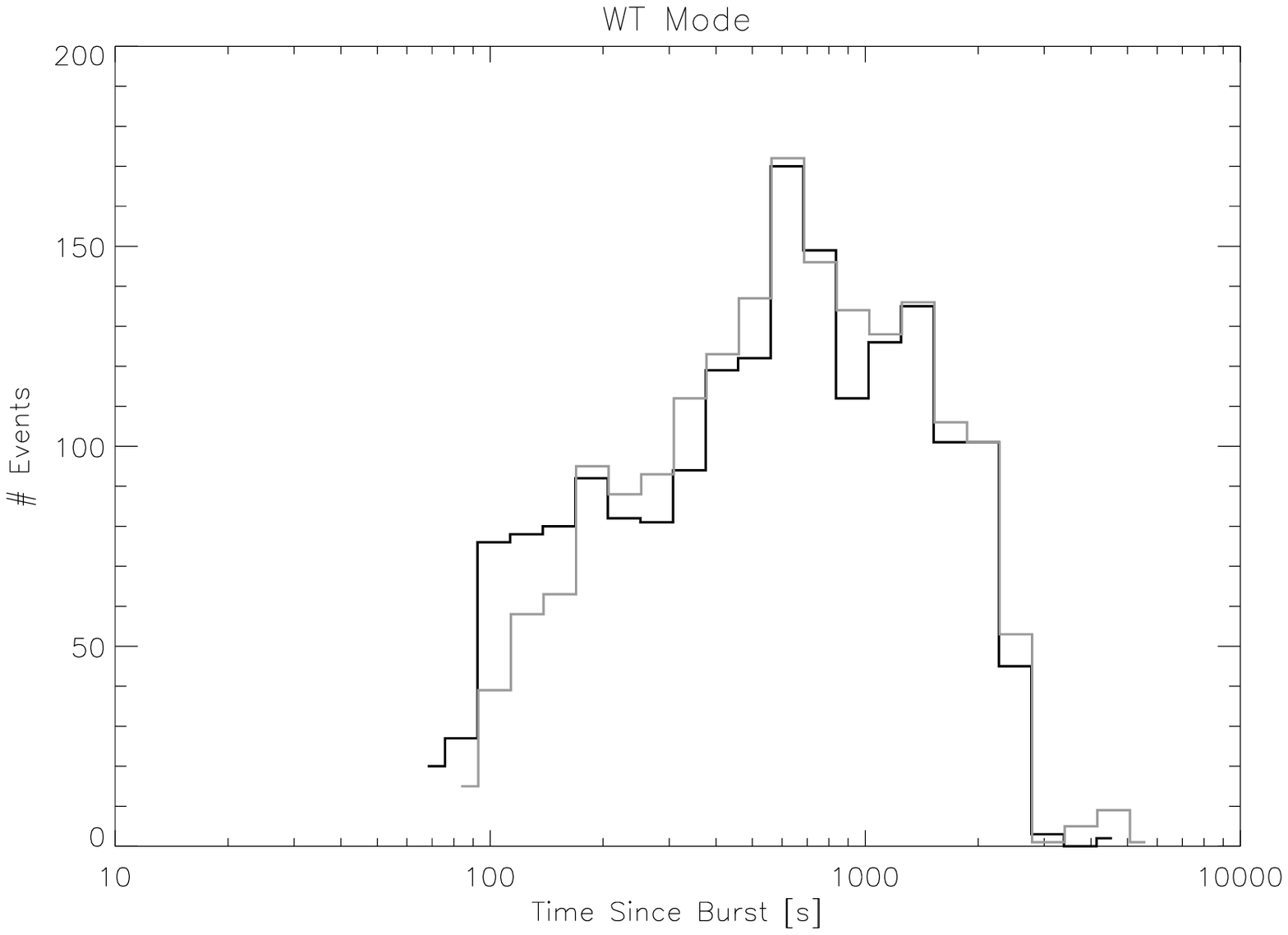}
\includegraphics[width=3.3in]{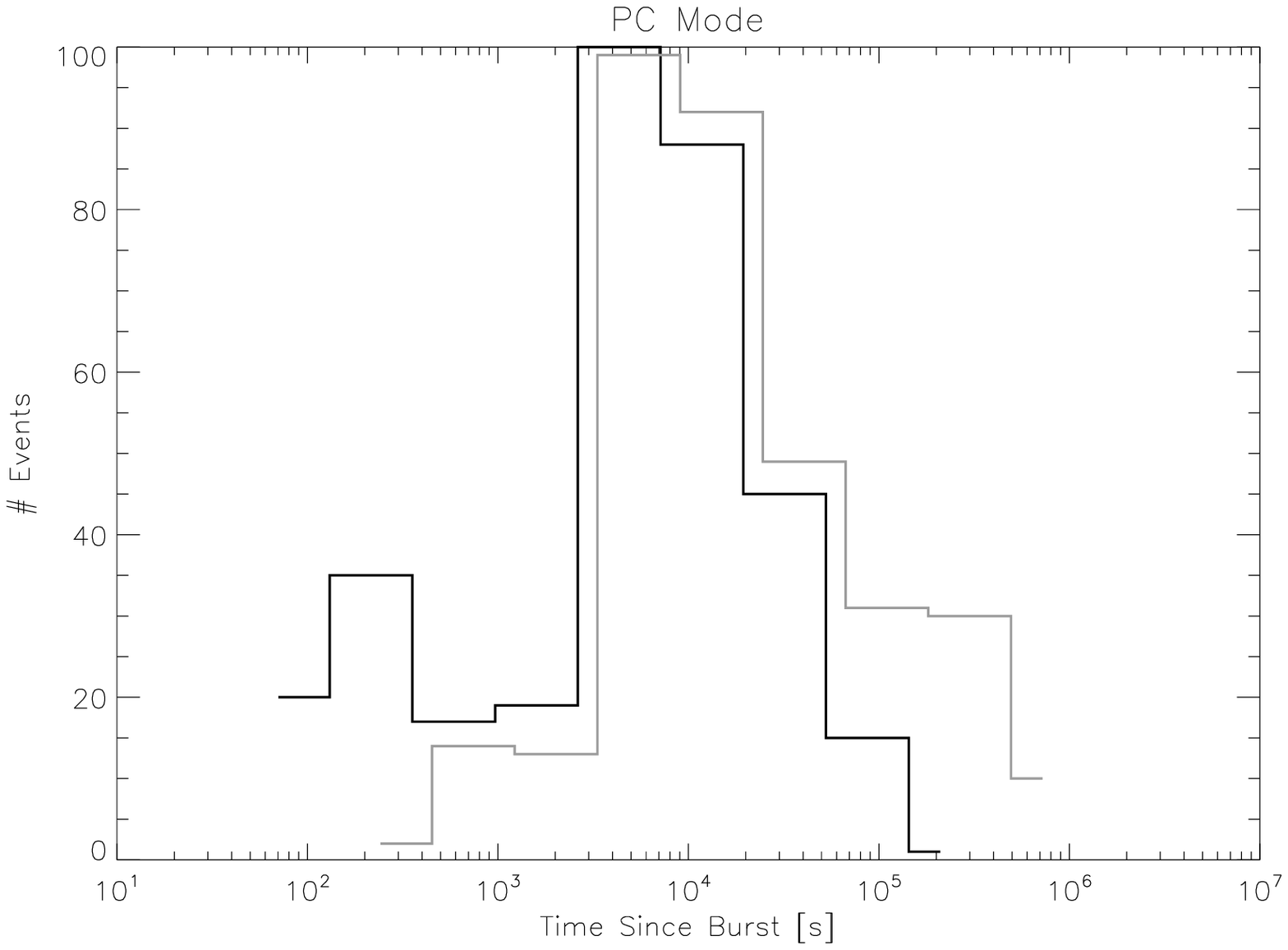}
\caption{\small (A) Start (black) and stop (grey) times for the WT mode
spectra
used in our analysis.  (B) Start and stop times for the PC mode data.
The duration
of the WT mode spectra range from 3.5 s to 4.5 ksec, with a median duration
of 17s.  The PC mode spectra durations range from 113 s to 2.4 Msec, with
a median duration of 9.9 ksec.
}
\label{fig:times}
\end{figure}

We fit the integrated spectra for each afterglow and 
also slice the data in time to search for transient emission features.
For the PC mode data, we consider data in the 0.3-10 keV energy range,
while we restrict to 0.5-10 keV for the WT mode data, where the low 
energy response appears to be poorly calibrated (see Section \ref{sec:cal}
for more details).
For the period between GRBs 050124 and 060313, we reduce data from 72 
PC-mode and 44 
WT-mode spectra, and sub-divide the PC and WT mode spectra into 208 and
944 spectra, respectively, each with $\sim 500$ counts.  The time 
coverage of the time-sliced spectra are plotted in Figure \ref{fig:times}.

\begin{figure}[H]
\includegraphics[width=3.5in]{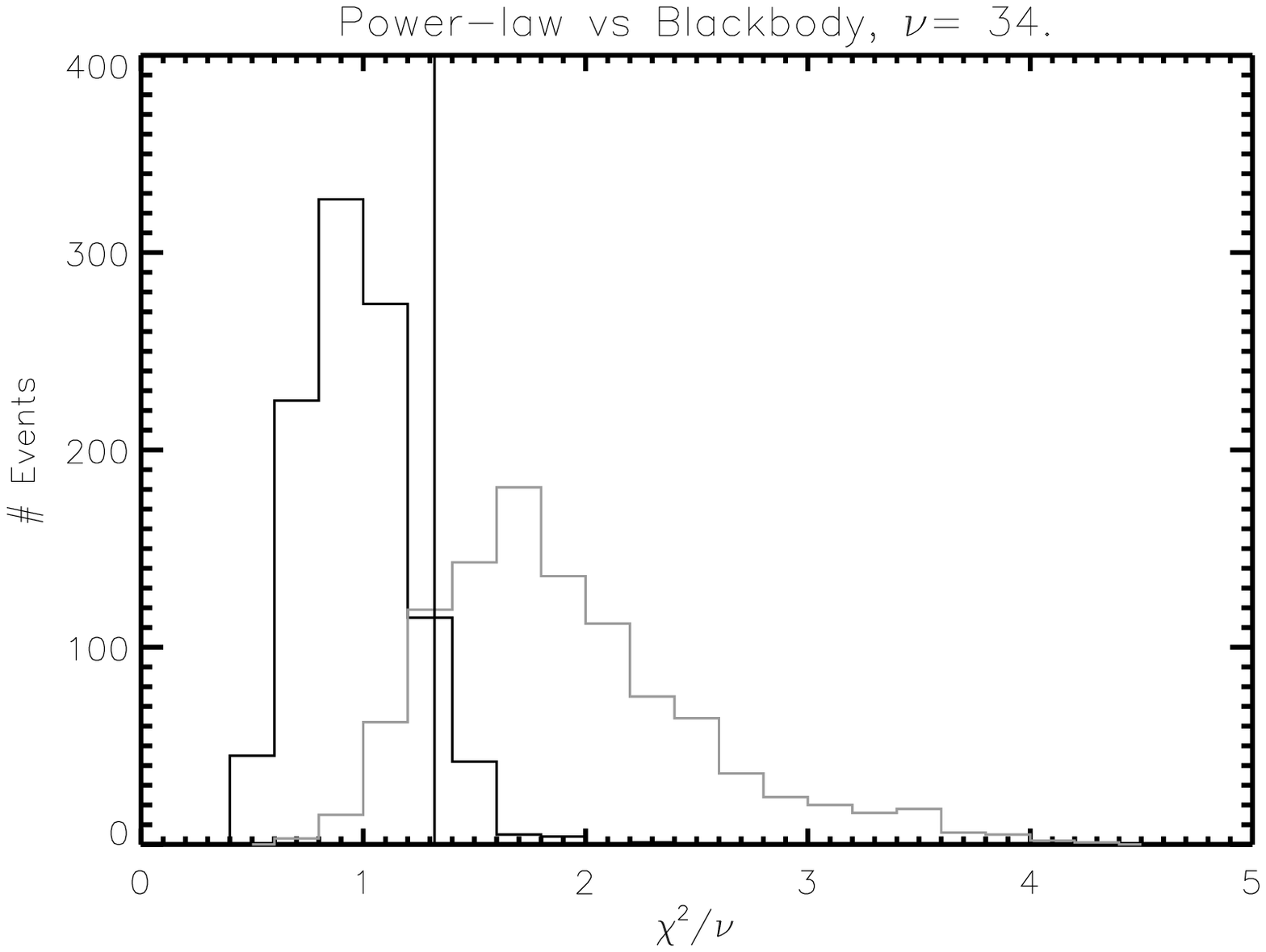}
\includegraphics[width=3.5in]{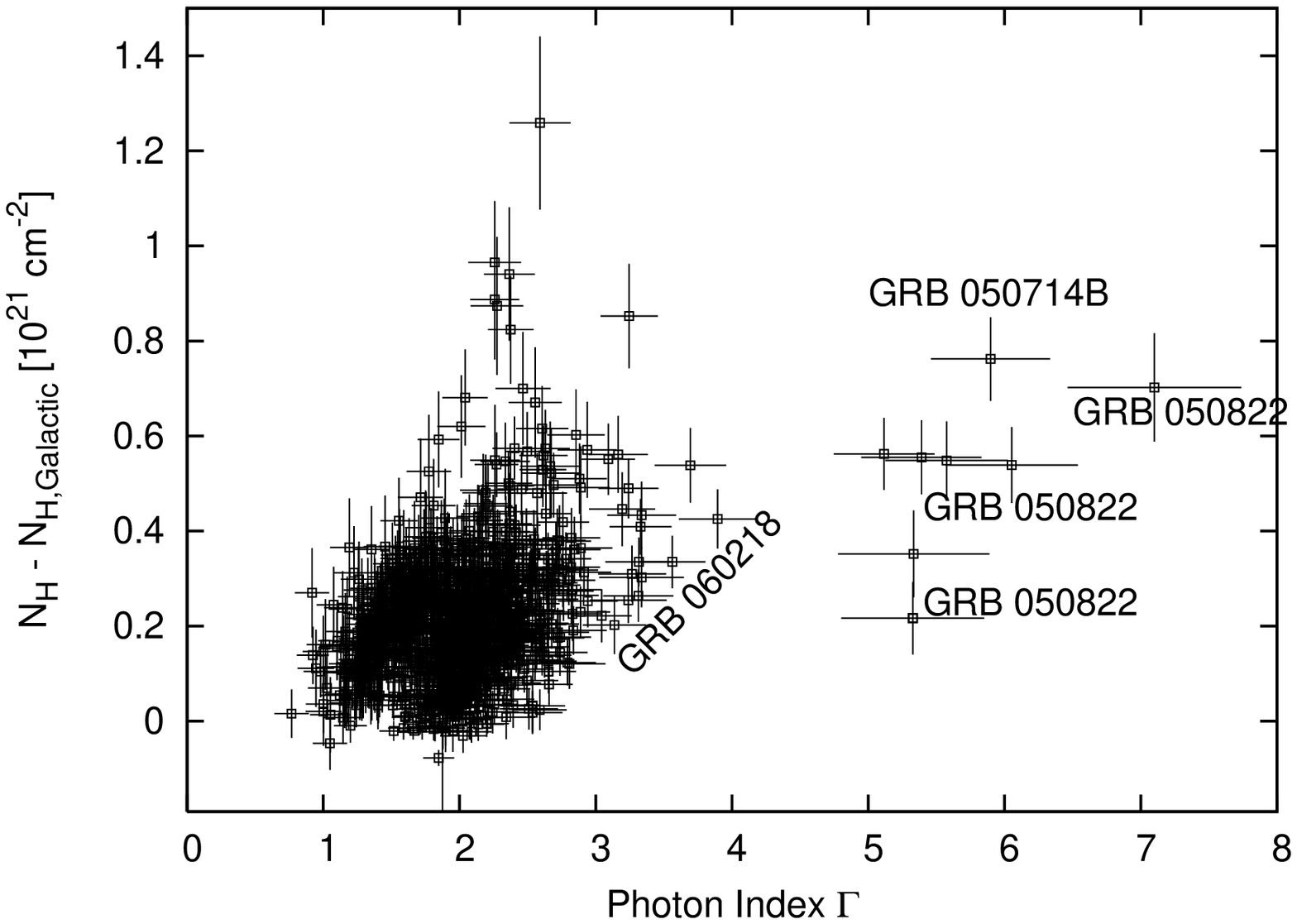}
\centerline{\includegraphics[width=3.5in]{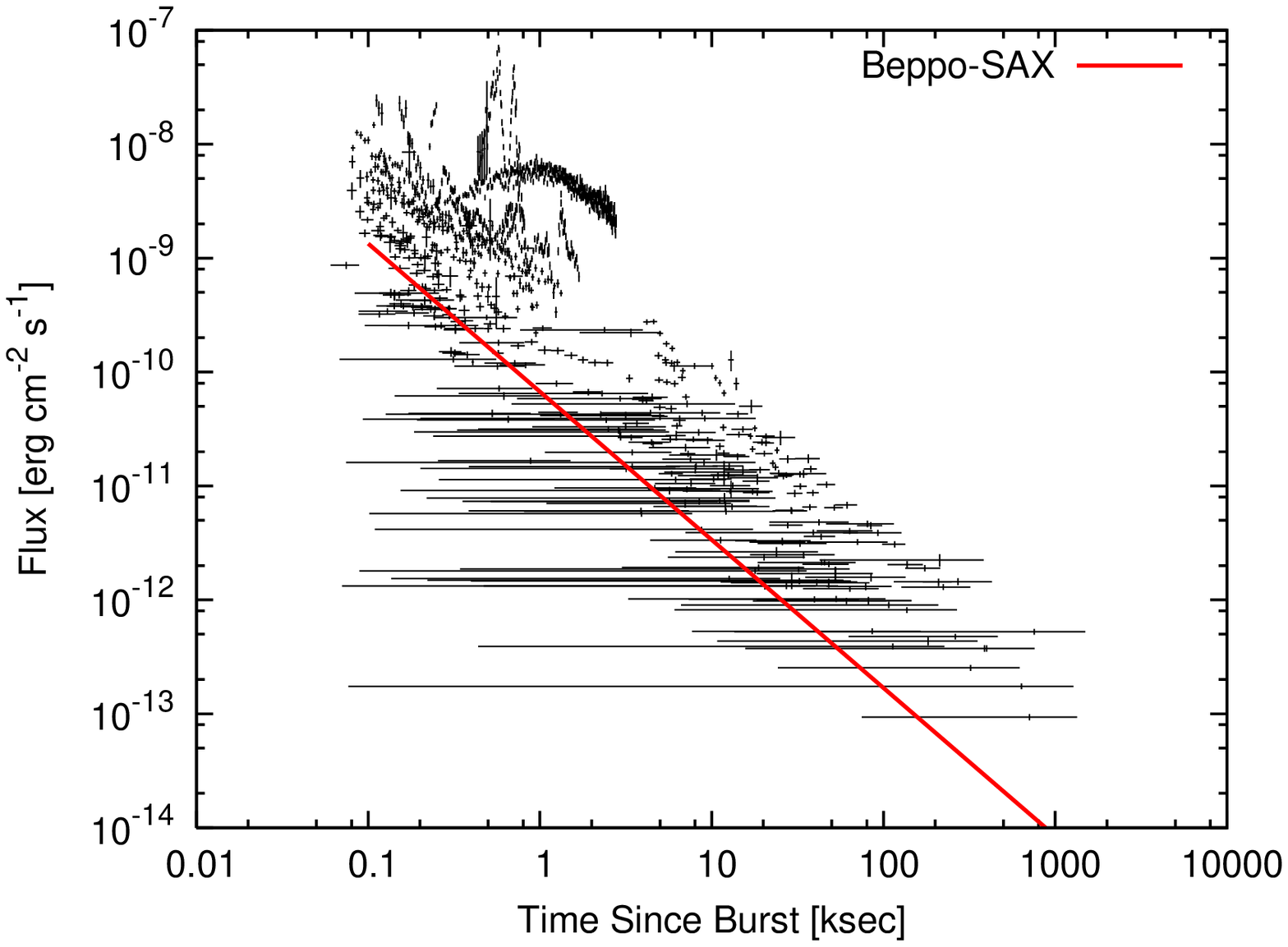}}
\caption{\small (A) Reduced $\chi^2$ values for the continuum fits, shifted
to a fiducial number of degrees of freedom $\nu=34$.  The
vertical line present the 90\% confidence rejection threshold, to
the right of which the fit should be rejected.  A small fraction of
the spectra ($\sim 10$\%) are acceptably fit with a blackbody model (grey line), however
most ($\sim 90$\%) are well fit by absorbed power-laws (black line). (B) Continuum
parameters $N_H$ and photon index $\Gamma$ for the absorbed power-law
model fits.  There are soft outliers (GRBs 050714B, 050822).
(C) Flux vs. time interval plot for the 500 count spectra.  Also, plotted is
the mean afterglow flux from {\it Beppo-SAX}~\citep{costa99}.  The {\it Beppo-SAX}~X-ray fluxes
are within $\pm 1$ dex of the red line.}
\label{fig:cont}
\end{figure}

We test the XRT data against two models for the continuum emission:
a power-law and a blackbody model, both with photoelectric absorption.
Overwhelmingly, the XRT data are well fit by absorbed power-laws.
There is a clear clustering of the photon indices $\Gamma$: the
median and median absolute deviation about the median are $\Gamma = 1.9 \pm
0.3$.  (Figure \ref{fig:cont}B).  The median column density
in excess of the Galactic value \citep{dickey1990} is
$N_H-N_{H,{\rm Galactic}} = (1.9 \pm 0.9) \times 10^{21}$ cm$^{-2}$,
marginally consistent with zero.  The integrated (i.e., not
time sliced) spectra show consistent values, $\Gamma = 1.9 \pm
0.2$ and $N_H-N_{H,{\rm Galactic}} = (0.9 \pm 1.2) \times 10^{21}$ cm$^{-2}$.
Soft outliers to these trends are labeled in Figure
\ref{fig:cont} and discussed more below.  The fluxes of the spectra cover
a broad range (Figure \ref{fig:cont}C).  They are roughly consistent with
fluxes previously measured for afterglows by {\it Beppo-SAX}~\citep{costa99}.

\section{The Emission Line Search: Efficiency and False Detection Rate}
\label{sec:search}

To search for lines in each spectrum, we fit
unresolved emission lines in addition to the power-law continuum in the 0.3-5.0 
keV band.  We allow for the possibility of multiple emission lines \citep[see,
e.g.,][]{reeves02}.  The lines are fitted successively, starting with
one line and adding a total of five lines.  The best-fit line location
at each step is found by convolving the fit residuals with the Line
Response Function (LRF) \citep[see, e.g.,][]{rutledge03}.  
The convolution tests for line energies on the $\delta E=5$ eV Ancillary Response 
File grid. 
The algorithm is described in more detail in \citet{butler05a}.  We model
the LRF versus PI energy bin as a Gaussian with an energy dependent 
width, $\sigma(E) = 29.4 (E/[1{\rm keV}])^{0.355}$ eV.  This functional 
form provides an
excellent fit to the core of the XRT LRF.  We estimate the significance
of each line set by applying the Likelihood Ratio Test (LRT) (also referred
to as the $\Delta \chi^2$ test) to the difference in observed $\chi^2$ values
prior to and after adding the emission line component.  These significance
estimates are then checked by comparing to the distribution of $\Delta
\chi^2$ produced by a Monte Carlo (MC) simulation (Section \ref{sec:search}).  We quote significance values
in units of normal distribution $\sigma$'s (i.e., a null hypothesis probability of
$2.7\times 10^{-3}$ is $3\sigma$, $5.7\times 10^{-7}$ is $5\sigma$, etc.).

For a spectrum containing $\sim 500$ counts, the typical 
energy binning oversamples the LRF full-width at half max by a factor $\sim 2$.
By fixing the number of counts in each spectra, it is a simple task to
determine MC line significances for the sample as a whole.
Such a determination for each spectrum would be prohibitively time
consuming.  To be sure that we are not missing features due to the 500
count constraint, we have also dyadically grouped (2,4,8, spectra together, etc.,
up to the full integration) and searched the spectra for each burst.
We impose a tighter trigger criteria (see Section \ref{sec:dyadic})
on these data.  In section
\ref{sec:others}, we discuss a line search for the integrated spectra of bright
($>10$ cts/s) XRT flares.

\begin{figure}[H]
\begin{centering}
\centerline{\rotatebox{270}{\includegraphics[width=3.5in]{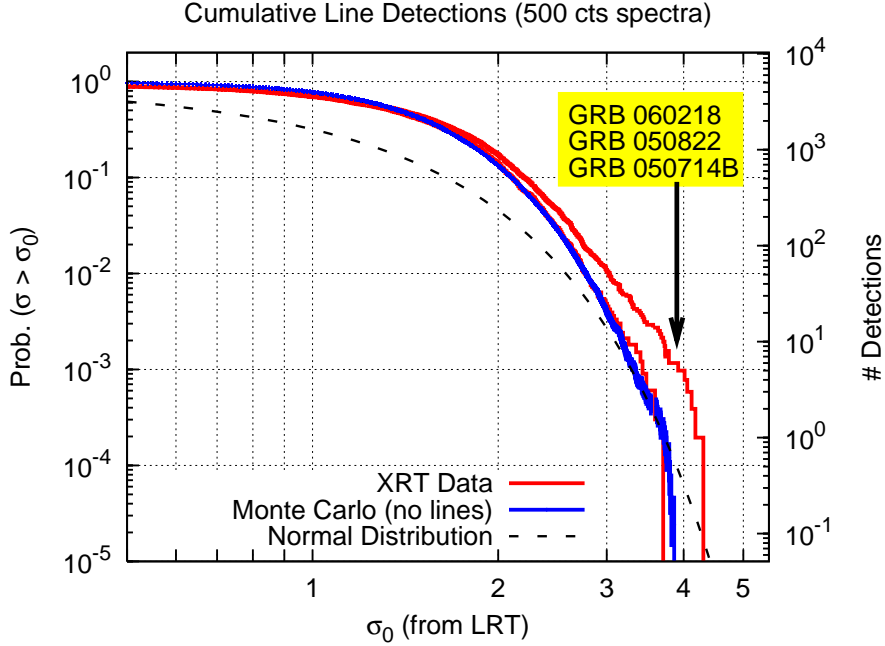}}}
\end{centering}
\caption{\small
The number of triggers (red lines) for 1-5 emission lines in XRT spectra
containing $\sim 500$ counts versus their estimated significance.  A
branch in the XRT curve is seen depending on whether or not 050714B, 050822,
and 060218 are included in the sample.  The bottom red curve does not
include these events and is roughly consistent (KS-test significance $\sim 3\sigma$)
with the Monte Carlo false event rate
(blue line).  Ignoring the data from the anomalously bright GRB~060218---which comprise 
$\sim 30$\% of the spectra under study---the excluded 2 bursts correspond to 4\% of the
the full sample spectra.  The Monte Carlo curve is determined from $10^5$ simulations
of a fiducial 500 count spectrum (Section \ref{sec:search}).  The dotted black line
is the expected false event for the LRT (i.e., the cumulative normal
distribution).}
\label{fig:multi_probs}
\end{figure}

Figure \ref{fig:multi_probs} displays our triggers for 1-5 emission
lines from the blind search through the full data set.
Searching 1152 spectra, the trials probability for finding one spectrum with a
$4.5\sigma$ trigger is 99\% confidence.  Given the sample median, lower and 
upper quartile values for the observed single-trial trigger significances found ---
$1.39\sigma$, $0.94\sigma$, and $1.83\sigma$, respectively, $4.5\sigma$ is also 
3 interquartile lengths 
from the upper quartile, which is a common measure for strong outliers \citep[e.g.,][]{devor95}.
We find a handful of triggers near or beyond this value, which occur in
the spectra of 3 bursts (e.g., Figure \ref{fig:multi_probs}).
As mentioned above, the significance estimates used here come from the LRT 
test.  We also check the validity of this procedure by simulating $10^5$ spectra
with 500 counts each and searching for lines in the simulated spectra.
For the simulation, we take $\Gamma=2$ and $N_H=10^{21}$
cm$^{-2}$ as representative values for the entire XRT sample.
The resulting (false) trigger rate for the most significant subset of 
1-5 emission lines from the search is plotted in blue
in Figure \ref{fig:multi_probs}.  There is a clear departure at low
significance from the model (dashed curve) for the LRT.  
Importantly, the departure is negligible and conservative for significances
$\gtrsim 3\sigma$.  
The validity and limitations of the LRT are discussed in detail
elsewhere \citep{yaqoob98,protassov02}.

\begin{figure}[H]
\begin{centering}
\centerline{\rotatebox{270}{\includegraphics[width=2.5in]{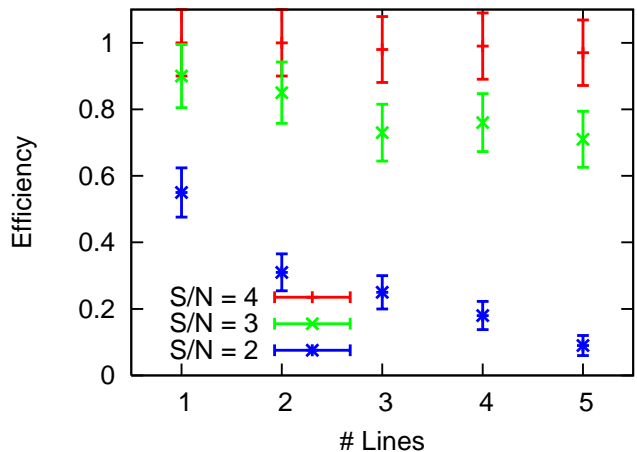}}}
\end{centering}
\caption{\small
The fraction ($=$ efficiency) of $10^3$ simulated spectra with lines of varying signal
to noise ($S/N$) where the trigger algorithm found the lines at
their true locations and with consistent $S/N$ values.  The efficiency
is near unity for 1-5 emission lines, each with $S/N\gtrsim 3$.  The
efficiency drops sharply for multiple lines with $S/N\lessim 2$.}
\label{fig:effic}
\end{figure}

We also perform simulations to determine the efficiency
of the algorithm for detecting lines known to be present.
Starting with the same continuum model used for the false trigger
tests, we add lines with $S/N=3$ (comparable to low $S/N$ values from lines
detected previously) randomly placed in energy over the 0.3-5.0 keV band.
Figure \ref{fig:effic} displays the efficiency (fraction
of iterations where the lines are detected at their true locations) of
the algorithm as the assumed line $S/N$ is varied in the MC.  The efficiency
drops sharply for $S/N\lessim 2$ but remains near unity for $S/N\gtrsim 3$.
We note that the trigger software autonomously finds \citep[see,][]{butler05a} each of the 
5 emission lines claimed for GRBs 011211 \citep{reeves02} and 030227 \citep{watson03}.
There is no clear dependence of the efficiency on the line energy, however,
the restricted energy band places clear limits on the chemical species to
which the search is sensitive.  For nearby events ($z\lessim 0.4$), the
upper energy cutoff of 5.0 keV results in non-detection of the Fe- group
elements.  The low-energy cutoff can inhibit detection of light element
lines (e.g., Ne at $z\gtrsim 2$, Si at $z\gtrsim 5$).
From the simulations, we estimate that LRT significances in $\sigma$
units are accurate to
$\sim 20$\% for $S/N\gtrsim 3$ lines with $\sigma_{\rm LRT} \gtrsim 3$.
Potential unresolved emission lines from astrophysically abundant species are detected with high efficiency
over a broad range of redshifts.   In principle, the search is also sensitive to broad emission
lines, which can be built up from the superposition of multiple narrow lines.
We now discuss 
the individual triggers in reverse chronological order.

\section{Line Emission Detections in Soft Portions of Three X-ray Afterglows}
\label{sec:bigthree}

\subsection{GRB~060218}

\begin{figure}[H]
\includegraphics[width=3.3in]{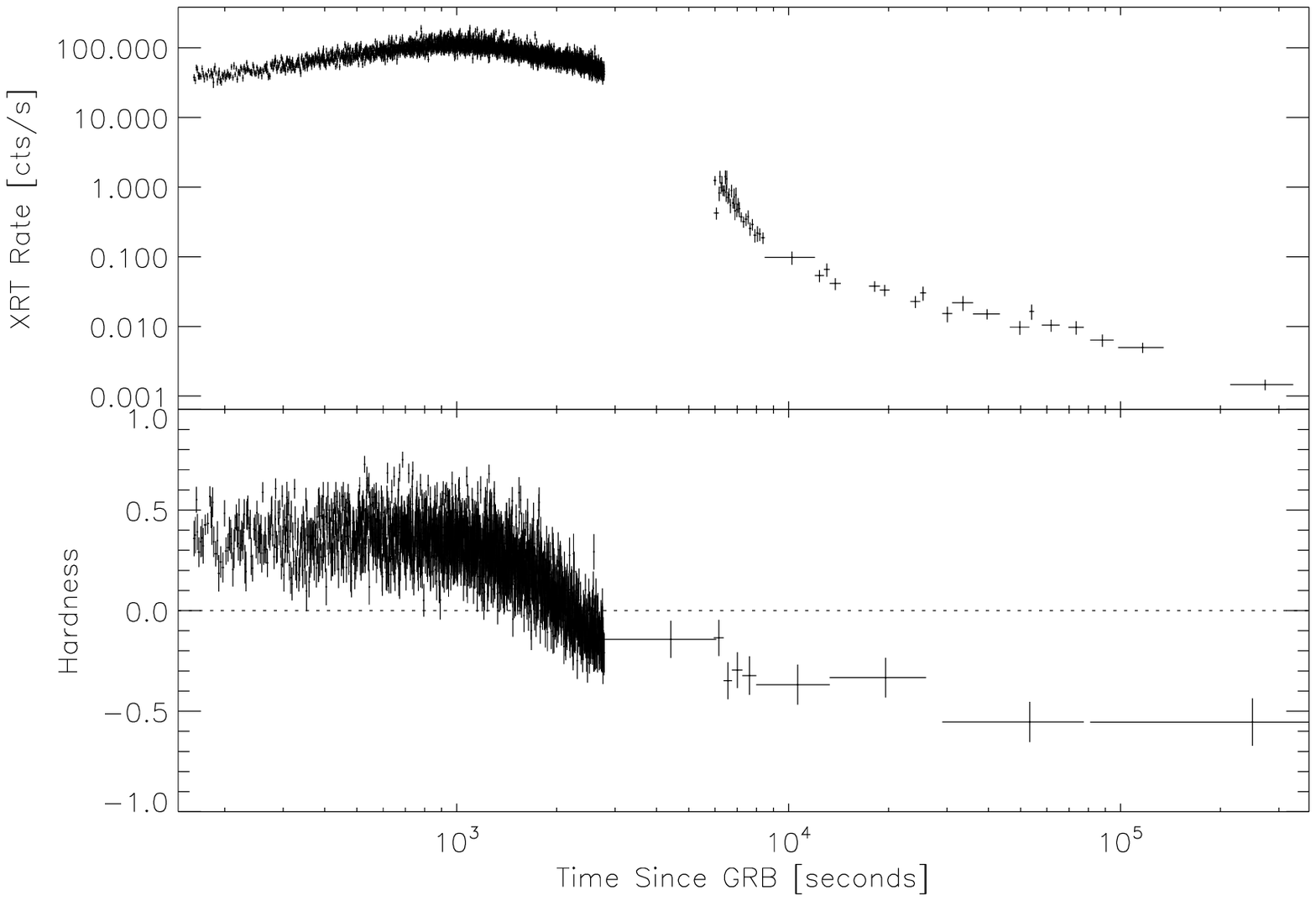}
\includegraphics[width=3.3in]{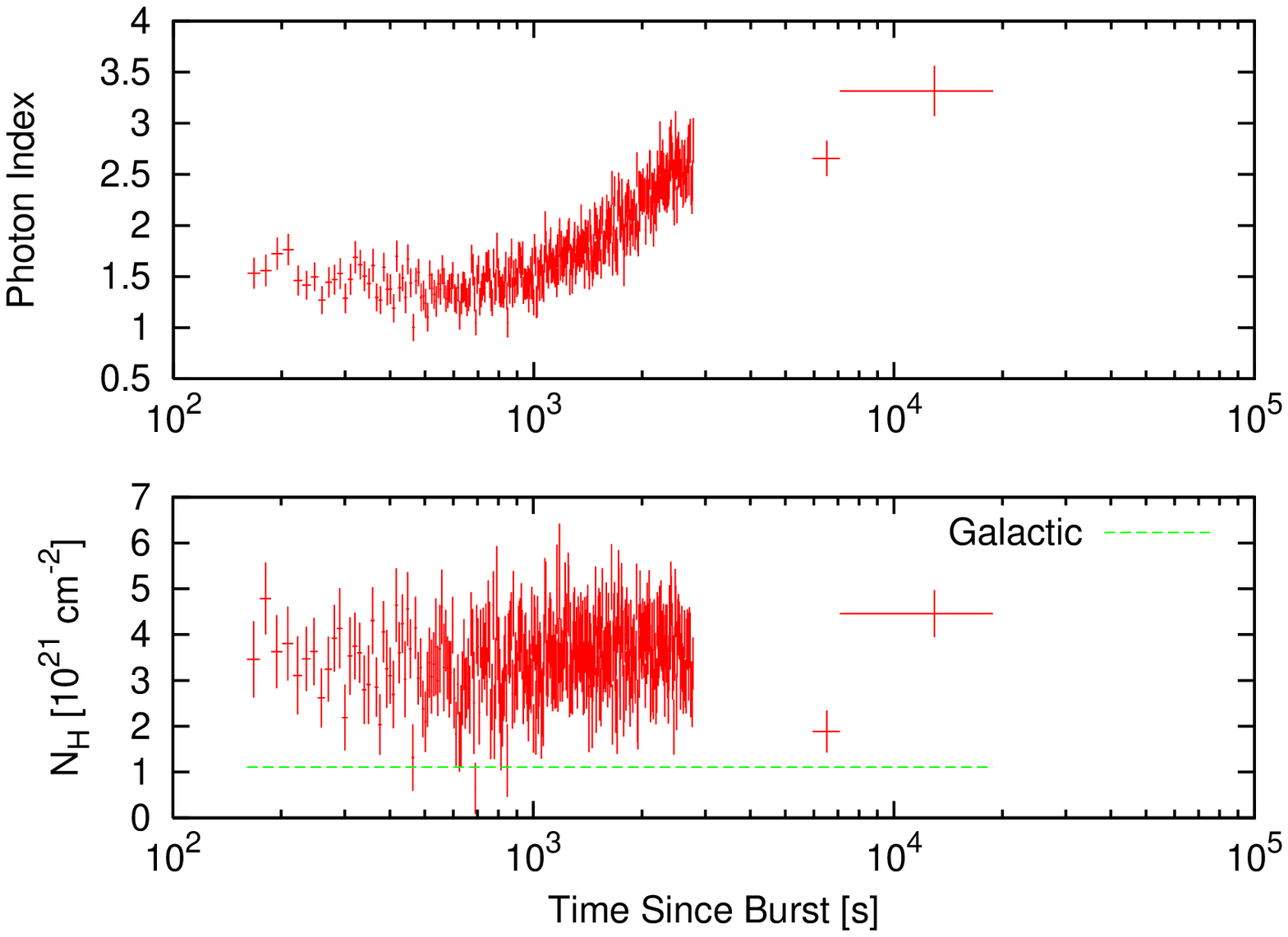}
\caption{\small
(A) The X-ray light curve for the GRB~060218 afterglow. ``Hardness'' is the difference
over the sum of the 1.5-8.0 keV counts and the 0.5-1.5 keV counts.
(B) Power-law fits to the afterglow.}
\label{fig:lc_060218}
\end{figure}

The flux from this event was extraordinarily high.  The light curve
displays a prominent hard to soft evolution during a broad
temporal rise, followed by a rapid decline and leveling off (Figure \ref{fig:lc_060218}).  
The soft portion of the afterglow (after $t\sim 1$ ksec) accounts for
much of the tail of the $\Gamma \sim 2-3$ distribution in Figure \ref{fig:cont}.
A preliminary spectral analysis is reported in \citet{cusumano06b}.
We group the WT (PC) mode counts into 355 (2) spectra, each containing
$\sim 500$ cts.  Due to the large data mass, we present our fit results in
plot form (Figure \ref{fig:lc_060218}B) for these finely resolved data,
providing tables below for the spectra with $\sim 16,000$ cts.  

\begin{figure}[H]
\centerline{\rotatebox{270}{\includegraphics[width=3.0in]{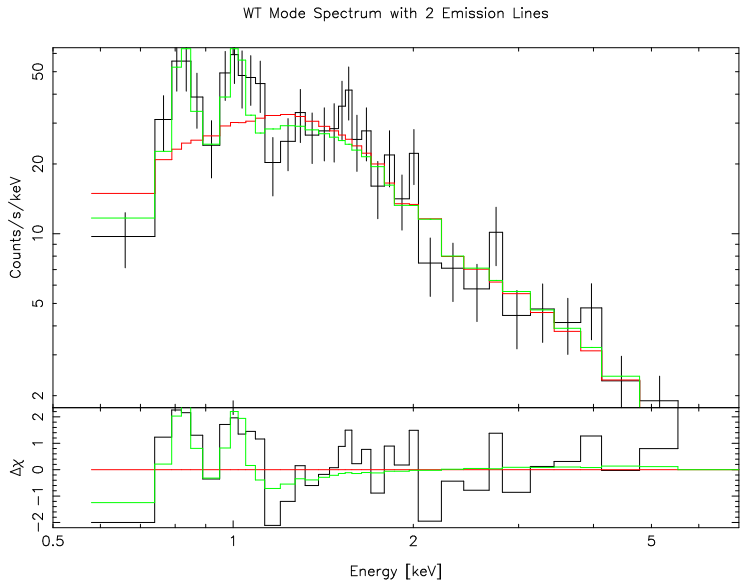}}}
\caption{\small
The XRT WT Mode spectrum from 2.317 to 2.326 ksec after GRB~060218.  Two
emission lines at 0.82 and 1.01 keV (green curve), possibly associated with L-shell transitions in Fe
at $z=0.033$ \citep{mirabal06}, improve the power-law fit (red curve) at $4\sigma$ significance.
In the residuals panel, the green curve shows the way the power-law residuals (black curve) are fit by the
power-law plus lines model.}
\label{fig:060218_linesB}
\end{figure}

Overall, the finely time-sliced data are well fit ($\chi^2/\nu=12149.5/12291$
for 356 spectra) with absorbed power-law's, with $N_H=(3.8 \pm 0.4)
\times 10^{21}$ cm$^{-2}$ on average \citep[consistently larger than the
Galactic value $N_{H,{\rm Galactic}}=1.11 \times 10^{21}$ cm$^{-2}$;][]{dickey1990}.
Toward the end of the WT mode data
at ($t=2.317-2.326$ ksec), the power-law fit becomes poor ($\chi^2/\nu=52.59/32$,
with $\Gamma=2.3\pm 0.2$, $N_H=(3.9\pm0.6) \times 10^{21}$ cm$^{-2}$, and unabsorbed flux
$f=(3.1\pm0.3) \times 10^{-9}$ erg cm$^{-2}$ s$^{-1}$ [0.5-10 keV]). The fit
is improved at $4\sigma$ significance ($\Delta \chi^2 = 24.69$, for 4 
additional degrees of freedom) with the addition of 2 emission lines (Figure \ref{fig:060218_linesB}).
The best fit line energies are 0.82$\pm$0.04 and $1.01\pm0.04$ keV.  
For redshifts near $z=0.033$ \citep{mirabal06}, the lines can be identified with
 K-shell transitions in H and He-like species of Ne or L-shell
 transitions from Fe-group elements.  The
equivalent widths and luminosities are 205$\pm$42 and 145$\pm$ 30 eV and
$3.6\pm0.2$ and ($1.9\pm 0.2) \times 10^{-10}$ erg cm$^{-2}$ s$^{-1}$,
respectively.  The blackbody model provides a poorer fit to the data in
this ($\chi^2/\nu=82.96/32$) and the other time regions.
A power-law plus blackbody model provides a mediocre fit ($\chi^2/\nu=42.94/30$).

As we discuss below, the integrated spectrum 
is also quite well fit by a thermal MEKAL \citep{mekal} plasma model ($\chi^2/\nu=26.90/30$)
with $kT=0.44\pm0.06$ and normalization (see note in Table 3) $3.0\pm1.3$ for solar abundances.
The 1-$\sigma$ lower limit on the abundance is 0.2 solar, with an undefined upper limit.
The line emission is dominated by L-shell Fe, 
with an Fe to Ne abundance ratio $1.5^{+22.5}_{-0.7}$, relative to solar.

\begin{figure}[H]
\centerline{\includegraphics[width=7.0in]{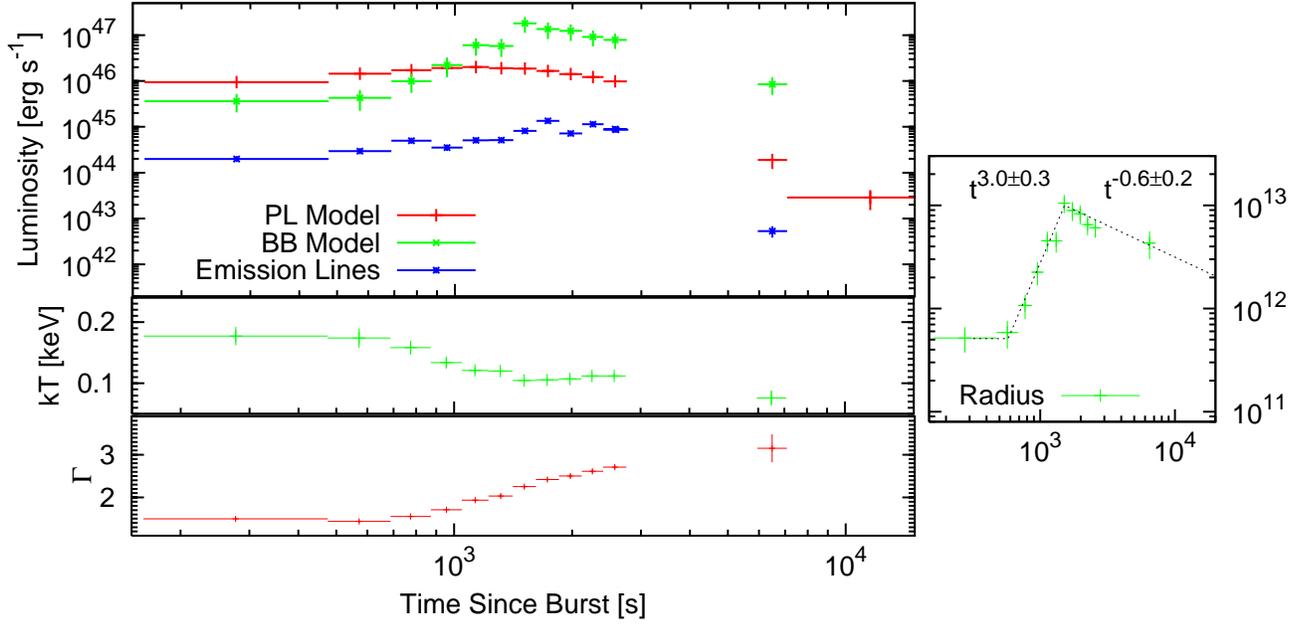}}
\caption{\small
A spectral deconvolution for GRB~060218 into soft and hard emission components.
The plotted power-law model components (Flux and photon index $\Gamma$) are for the
power-law plus blackbody model. The blackbody radius is measured in cm.}
\label{fig:lc_060218_2}
\end{figure}

There is strong evidence that the soft emission component is present throughout
an extended portion of the observation.
Only $\sim 50$\% of the dyadically grouped spectra containing $>500$ cts 
are found to
be adequately fit by the simple absorbed power-law models.  Particularly
for the spectra containing 16,000 or more counts, a soft component near
1 keV is strongly required (power-law model alone rejected at $>90$\% confidence
in 11 of 12 spectra; Table 1).
With the inclusion of the blackbody component or emission lines in addition to the
power-law, each fit to the $\sim $16,000 cts spectra yields $\chi^2/\nu\lessim 1$.

Figure \ref{fig:lc_060218_2} shows the results of the spectral
deconvolution of these data into power-law and soft components.
The power-law component displays a smooth hard to soft evolution in time.
In the case of the emission line model, the lines in the 0.8-1.0 keV energy
range, are detected in multiple time intervals (Table 2), 
with  high ($>5\sigma$) significance triggers coming after $t\sim 1$ ksec.
In addition to the lines at $E\lessim 1$ keV discussed above, triggers in the 3-4 keV range are found
(Table 2),
possibly associated with H- or He-like Ar and Ca.  

The soft-component 
luminosity $L$ in both cases appears to peak after the power-law component.
The blackbody temperature declines mildly after $t=580$s as $(0.19\pm0.01) (t/580{\rm s})^{(-0.54\pm0.08)}$ keV
to $kT=0.105\pm0.004$ keV at $t=1.52$ ksec, then remains constant and possibly declines at late time.  
Using the standard formula for the blackbody radius:
\begin{equation}
\label{equation:bbody}
  R_{\rm BB} = 2.78 \left({L \over 10^{46} {\rm erg~s}^{-1}}\right)^{1/2} \left({kT \over 0.1 {\rm ~keV}}\right)^{-2} \times 10^{12} {\rm cm},
\end{equation}
and $z=0.033$ \citep{mirabal06},
the blackbody radius increases, peaks, then decreases with the time dependences
shown in Figure \ref{fig:lc_060218_2}. 
Here and throughout, we assume a cosmology with (h,$\Omega_m$,$\Omega_{\Lambda}$)$=$(0.65,0.3,0.7).
We find consistent behavior independent of whether or not we fix $N_H$ or $kT$ in the
fits.  
Consistent values for the initial blackbody radius are reported by \citep{campana06}.  However, they find that the
radius continues to grow in time during the XRT observation without peaking, a scenario that
we find to be ruled out at the $\sim 7\sigma$ level.  

\begin{figure}[H]
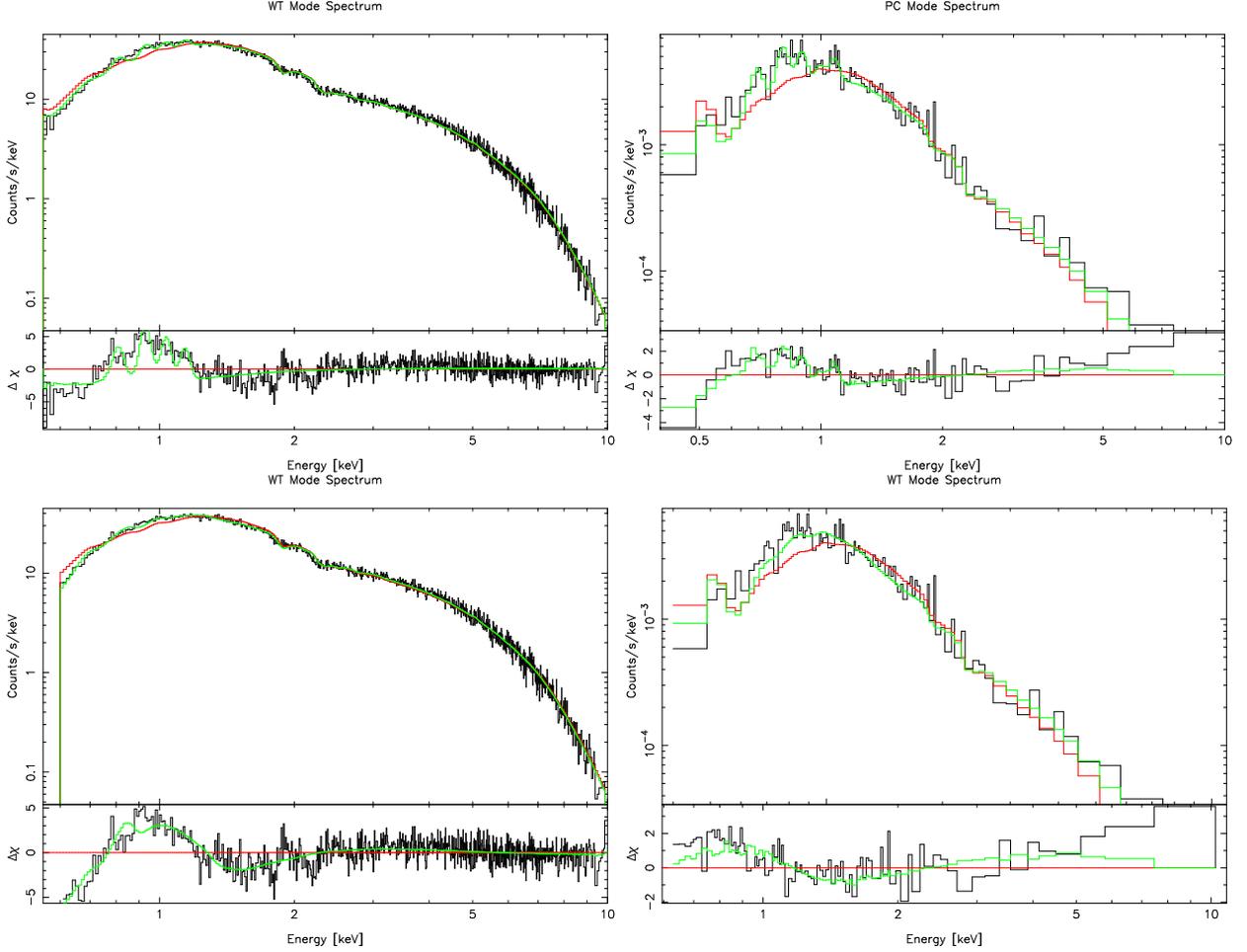

\rotatebox{270}{\includegraphics[width=2.5in]{f8a.ps}}
\rotatebox{270}{\includegraphics[width=2.5in]{f8b.ps}}
\rotatebox{270}{\includegraphics[width=2.5in]{f8c.ps}}
\rotatebox{270}{\includegraphics[width=2.5in]{f8d.ps}}
\caption{\small
Power-law models (red curves) fail to describe the soft emission in the integrated WT mode ($t=160$s-2.8ksec)
and PC mode ($t=5.9$ksec-3.1Msec) spectra for GRB~060218 (Table 3).
(A) Top two plots.
Both data sets are better fit with the addition of 4-5 emission lines (green curves,
also plotted for the power-law model residuals):
WT mode, 4 lines, $20.6\sigma$, $\Delta \chi^2=460.03$, for 8 additional degrees of freedom;
PC mode, 5 lines, $6.9\sigma$, $\Delta \chi^2=75.37$, for 10 additional degrees of freedom.
(B) Bottom two plots.  The fits to both data sets are also markedly improved with the addition
of a blackbody continuum component (Table 3).
}
\label{fig:060218_whole}
\end{figure}

Figure \ref{fig:060218_whole} shows
the integrated WT and PC mode spectrum for the afterglow.  Continuum spectral fits are reported
in Table 3.
Prominent residuals are present in both spectra 
near 1 keV.  The WT and PC mode data
can be fit with 4 emission lines at energies 0.80 $\pm$ 0.03,
$0.9 \pm 0.03$, 1.00 $\pm$ 0.03, and 1.1 $\pm$ 0.03 keV in addition to the power-law
continuum.  
The WT mode trigger is the most significant trigger in the entire sample
by a large margin (see Section \ref{sec:dyadic}).
The line luminosities are $\approx 2 \times 10^{-11}$ erg cm$^{-2}$ s$^{-1}$ in the WT mode spectrum
and $\sim 10^3$ times fainter in the PC mode data.  Oppositely, the equivalent widths increase
by a factor $\sim 5$ from $\approx 20$ eV in the WT mode data, indicating that the line fluxes
decrease less slowly than the continuum flux (see, also, Figure \ref{fig:eqwid}).
If we relax the upper limit on the line energy search,
there is a weak ($\sim 2\sigma$) trigger on a possible K-shell Fe- series line at $E=6\pm1$ keV
in the PC mode spectrum, with
an equivalent width of order 1 keV (e.g., residuals in Figure \ref{fig:060218_whole}B).
The further addition of lines to either spectrum does not significantly improve the
fits.  

The poor apparent $\chi^2/\nu=1280.53/760$ in the case of the WT mode data
probably should not be taken too seriously,
because the flux is high and the inclusion of a small $\lessim 5$\% systematic error component
leaves the fit statistically acceptable.  The spectra can also be fit with a thermal plasma
component (assuming solar abundances)
in addition to the power-law (Table 3)--which reinforces the
notion that some of the soft emission may be in the form of discrete lines--or with
a blackbody in addition to the power-law.

Comparing the hypothesis of emission lines to that of a blackbody (in 
additional to a non-thermal continuum) for GRB~060218, we find that the
event is a particularly strong line candidate.  The power-law plus two line
model is an excellent fit to the 500 cts spectrum discussed above 
($\chi^2/\nu=27.9/28$), whereas the power-law plus blackbody model fit
is mediocre ($\chi^2/\nu=42.94/30$.)

\subsection{GRB~050822}

\begin{figure}[H]
\includegraphics[width=3.3in]{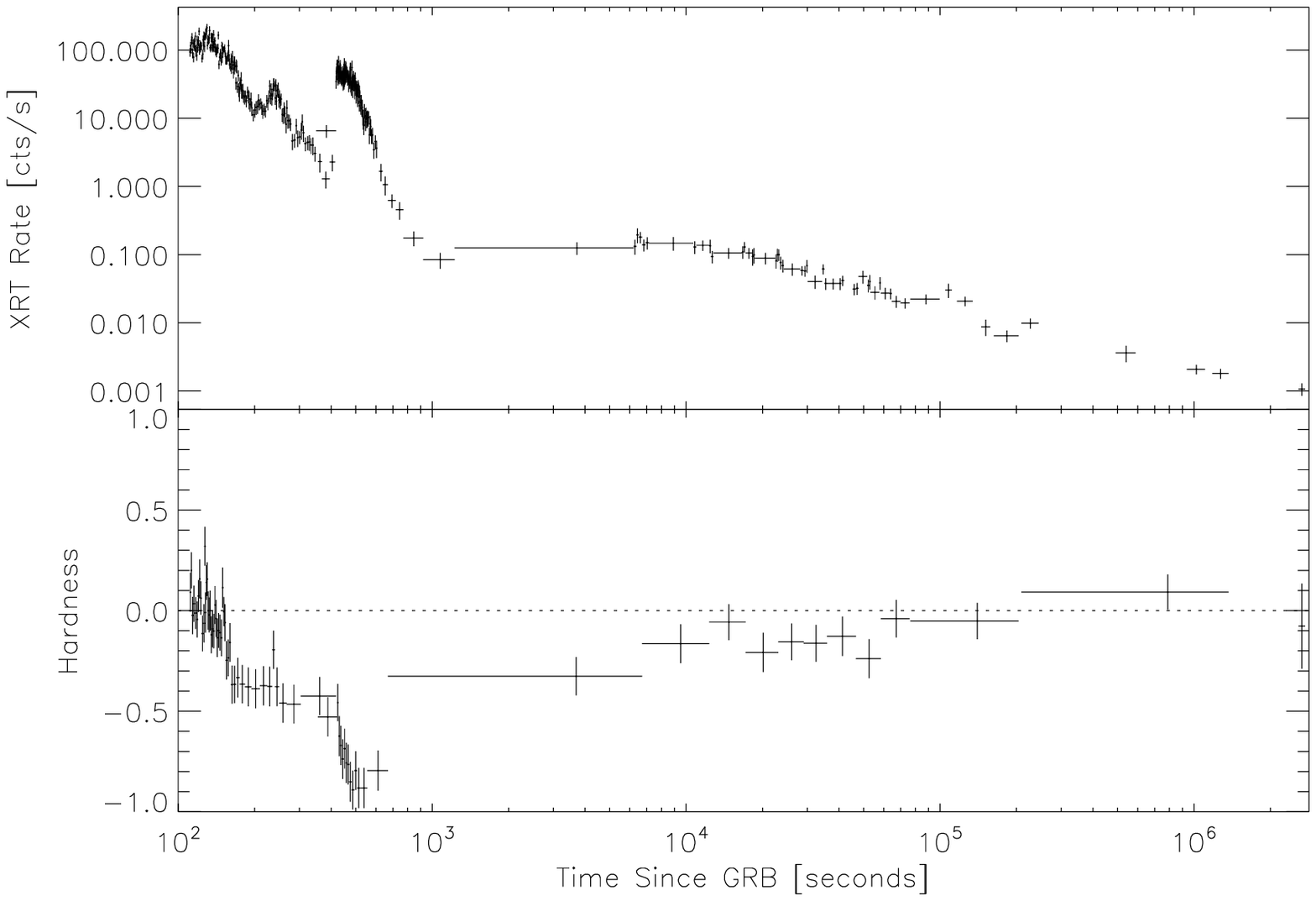}
\includegraphics[width=3.3in]{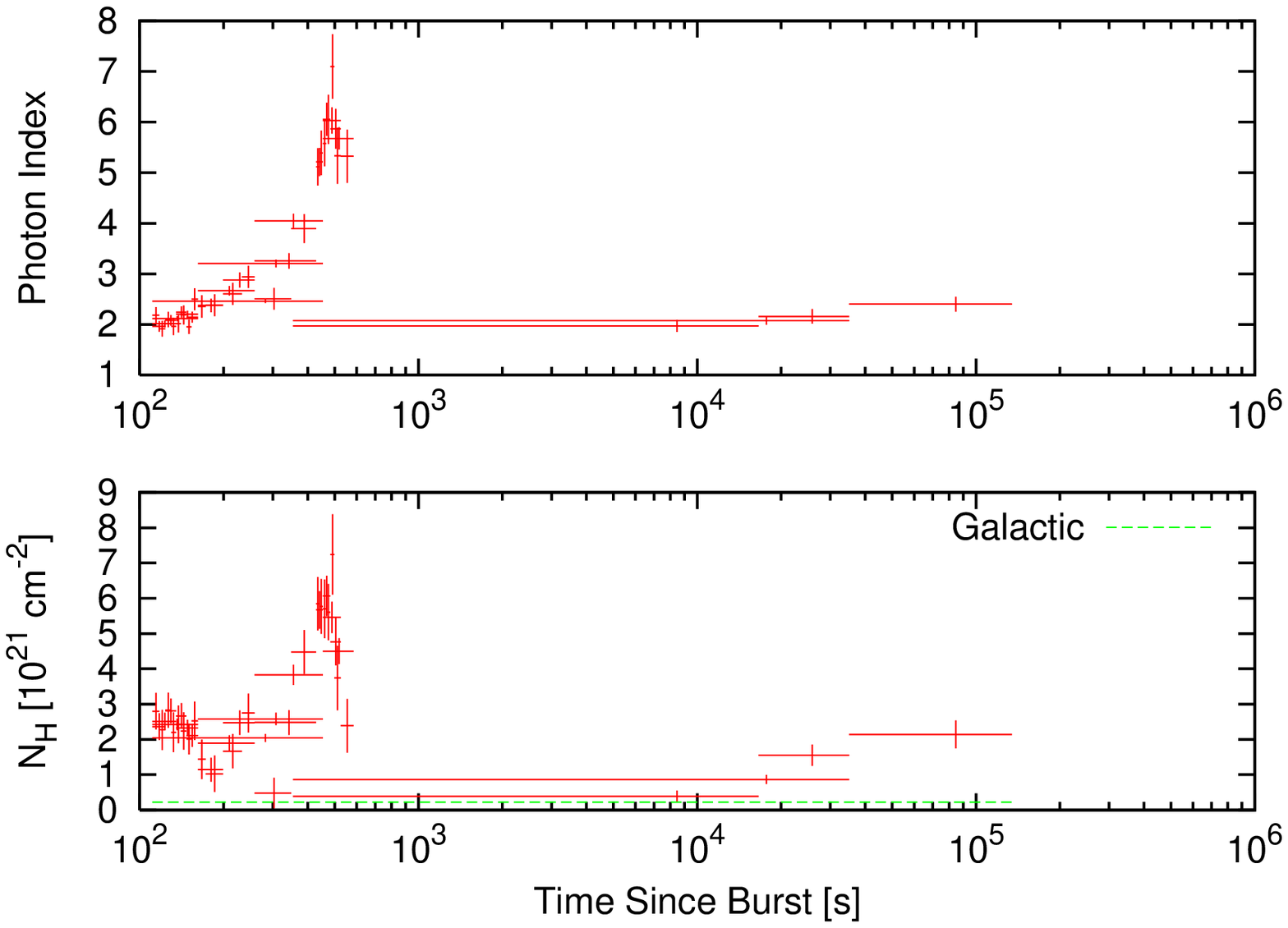}
\caption{\small
(A) The X-ray light curve for the GRB~050822 afterglow.  There is a
strong flare at $t\sim 500$s, during which the spectrum becomes very soft.
(B) Spectral evolution for GRB~050822. A dramatic increase in the spectral curvature near
400s is clear from a jump in the photon index and best-fit absorption column.}
\label{fig:050822_lc}
\end{figure}

This event is a clear outlier in Figure \ref{fig:cont}.  A preliminary spectral analysis has
been reported by \citet{godet05}.  The spectrum
displays a gradual softening in time which increases abruptly with
the impulsive flare at $t\approx 430$s (Figure \ref{fig:050822_lc}B).
The continuum spectrum during the flare (430s to 550s after the burst)
is well-fit by a blackbody
(Table 4), with a decreasing temperature in time
$kT = 0.26-0.17$ keV (observer frame).  There is an indication that the blackbody radius 
from Equation \ref{equation:bbody} may increase in time
by $\sim 50$\%, starting at $R_{\rm BB}= (2.7\pm0.5)\times 10^{12}$ cm at 
$t=430s$ and peaking at $R_{\rm BB}=(4.5\pm1.3) \times 10^{12}$ cm at $t=490$s, 
assuming $z=1.2$ (see below).

\begin{figure}[H]
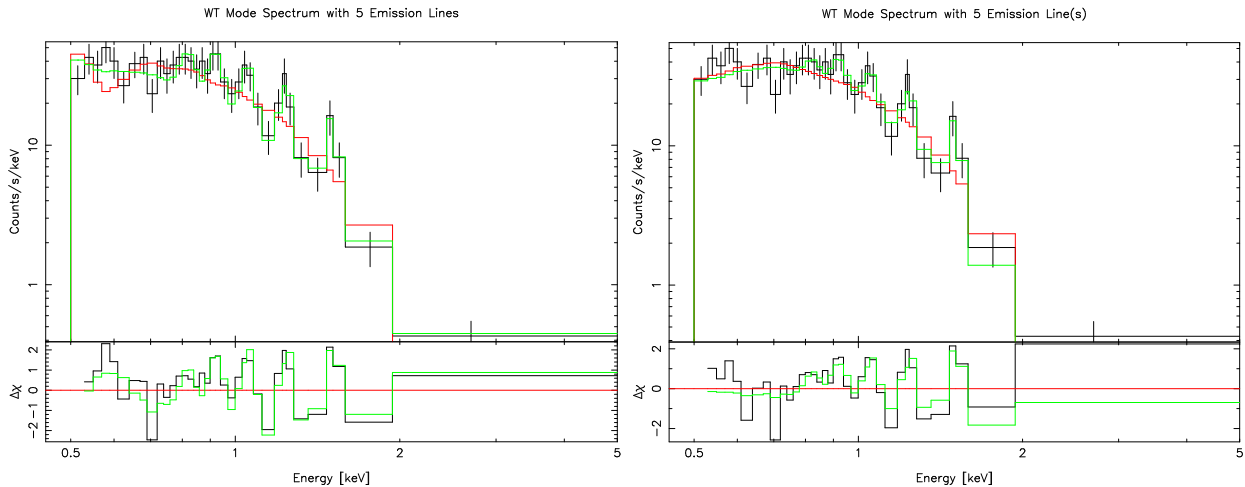

\rotatebox{270}{\includegraphics[width=2.5in]{f10a.ps}}
\rotatebox{270}{\includegraphics[width=2.5in]{f10b.ps}}
\caption{\small
(A) In interval \#14 during the GRB~050822 flare, emission lines at 0.81, 0.91, 1.04, 1.23,
and 1.49 keV (green curve), possibly associated with Si~XIII, Si~XIV, S~XVI, S~XV, and Ar~XVII at $z=1.2$, improve
the power-law model fit (red curve) at $4.4\sigma$ significance (Table 5).
(B) The same but for a blackbody continuum model.  The lines are $2.8\sigma$ significant.
For the residuals panel in each plot, the green curves show the way the continuum model residuals (black curves) are fit by the
continuum plus lines model.}
\label{fig:050822_flare}
\end{figure}

A power-law model also provides an acceptable fit 
to the data during the flare, except during the penultimate time 
interval, 489.5 to 509.4s (interval \#14 in Table 4; $\chi^2/\nu$=54.71/33, rejectable
at 99\% confidence).  The blackbody model fit is marginally better
in interval \#14 ($\chi^2/\nu$=51.06/33, rejectable at 98\% confidence)
than the power-law fit.
An added set of 5 emission lines improves the power-law fit during interval \#14
at $4.4\sigma$ significance ($\Delta \chi^2 = 40.56$, $\nu=10$;
Table 4).  The same 5 lines
improve the blackbody fit at $2.8\sigma$ significance ($\Delta \chi^2 = 25.03$, 
$\nu=10$).  The continuum fit residuals are narrow (Figure \ref{fig:050822_flare}); 
neither the power-law nor the blackbody fit for interval \#14 is improved by adding 
an additional blackbody or power-law continuum component.

We test for the
presence of the line set found for interval \#14 by fitting the best the line set
model to the data from the other time intervals.
We allow one floating normalization, with the other line parameters fixed relative
to this normalization.
If the line-set from interval \#14 is present during the flare outside of interval \#14
(i.e., in intervals 11, 12, 13, and 15), the combined flux from the 5-line set must be lower
by a factor $>4$ (90\% confidence).  Thus, the overall or relative flux of each line
appears to be changing in time.  If we take the $\sim 20$s period during which the line
triggers are found to be most sigificant as representative of the period of active
line emission, then the possible detection appears very unusual and probably not credible given
the models for line emission (Section \ref{sec:mech} below).
We note that there is a weak trigger on the line set at earlier times (intervals \#3-\#5), with a mean
flux consistent with that found for interval \#14 but also consistent with zero at the
1-$\sigma$ level.  

By hand, we combine the PC and WT mode spectra in the tail of the
flare (see Figure \ref{fig:050822_flare}) and use the PC mode redistribution matrix to
perform a fit.  The data in this interval (\#19 in Tables 4
and 5) are also poorly fit by either continuum model and show modest
significance evidence for 5 emission lines.   Thus, the very short duration of the
line trigger may be due to the detection threshold and may not reflect the physical emission timescale.

The redshift is currently unknown for this event.
The line centroids during interval \#14 (Table 4) are too
closely spaced to allow for an association with H-like ions from
light metals only \citep[e.g.,][]{reeves02,watson03}.  However,
if we allow the He-like species, identifications are possible and
non-unique.  One possible association for the 5 lines in Figure
\ref{fig:050822_flare}B is: Ar~XVII, S~XVI, S~XV, Si~XIV, and
Si~XIII at $z=1.2$.  From Table 5, the lines at
$E= 0.81$, 0.91, 1.04, and 1.23 keV are detected in multiple time intervals.
A line near $3.5$ keV is detected in two time intervals and could
be associated with H-like Co or Ni at $z=1.2$.
A power-law plus MEKAL
model provides a poor fit due to the closely spaced emission lines.

There is one minor caveat relevant to our analysis of this event.
We include two detector columns (RAW X columns \#290 and \#291) in the WT mode data which are normally
discarded in the standard {\tt xrtpipeline} processing.  These are neighboring 
columns (sections of which have been found to behave anomalously) to columns which were damaged as the result of a possible micro-meteorite
impact\footnote{http://swift.gsfc.nasa.gov/docs/swift/analysis/xrt\_digest.html}.
We find that the continuum spectral fit parameters are consistent whether the
columns are retained or rejected.  However, because the two columns are near the center
of the source extraction region for a large part of the observation, the loss in source
flux associated with rejecting the columns mildly reduces the line trigger significance to $3.5\sigma$.
Here the time extraction is broadened to 480-512s, in order to get 510 counts in the spectrum.

Comparing the hypothesis of emission lines to that of a blackbody (in
additional to a non-thermal continuum) for GRB~050822, we find that this
event, like GRB~060218, is a strong line candidate.  The power-law plus five line
model provides an excellent fit, ($\chi^2/\nu=14.15/23$), whereas a power-law plus blackbody improves the power-law fit
very little.  We note that the line sigificance degrades (from $4.4\sigma$ to $2.8\sigma$) if
the underlying continuum is assumed to be a blackbody.

\subsection{GRB~050714B}

\begin{figure}[H]
\includegraphics[width=3.3in]{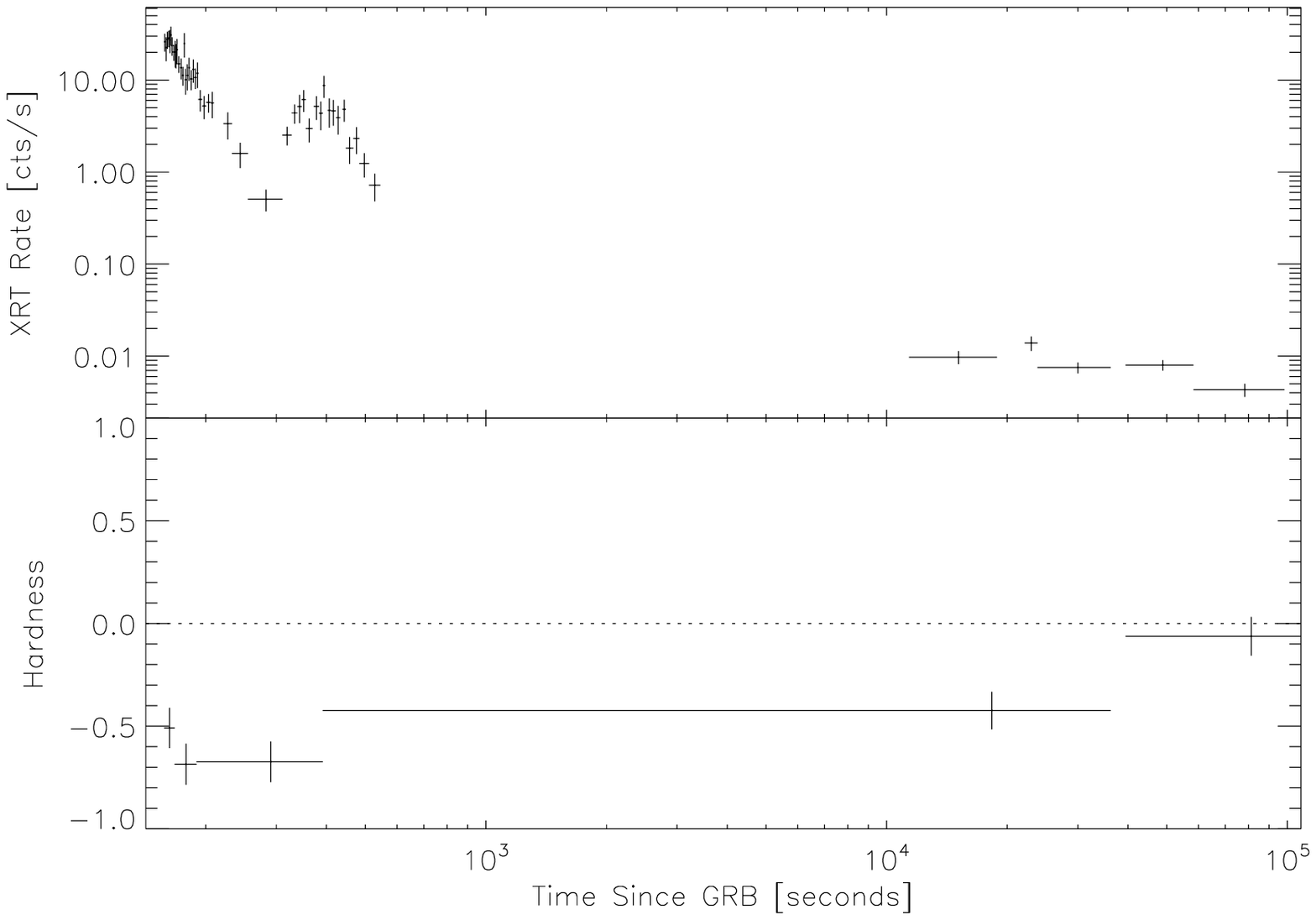}
\includegraphics[width=3.3in]{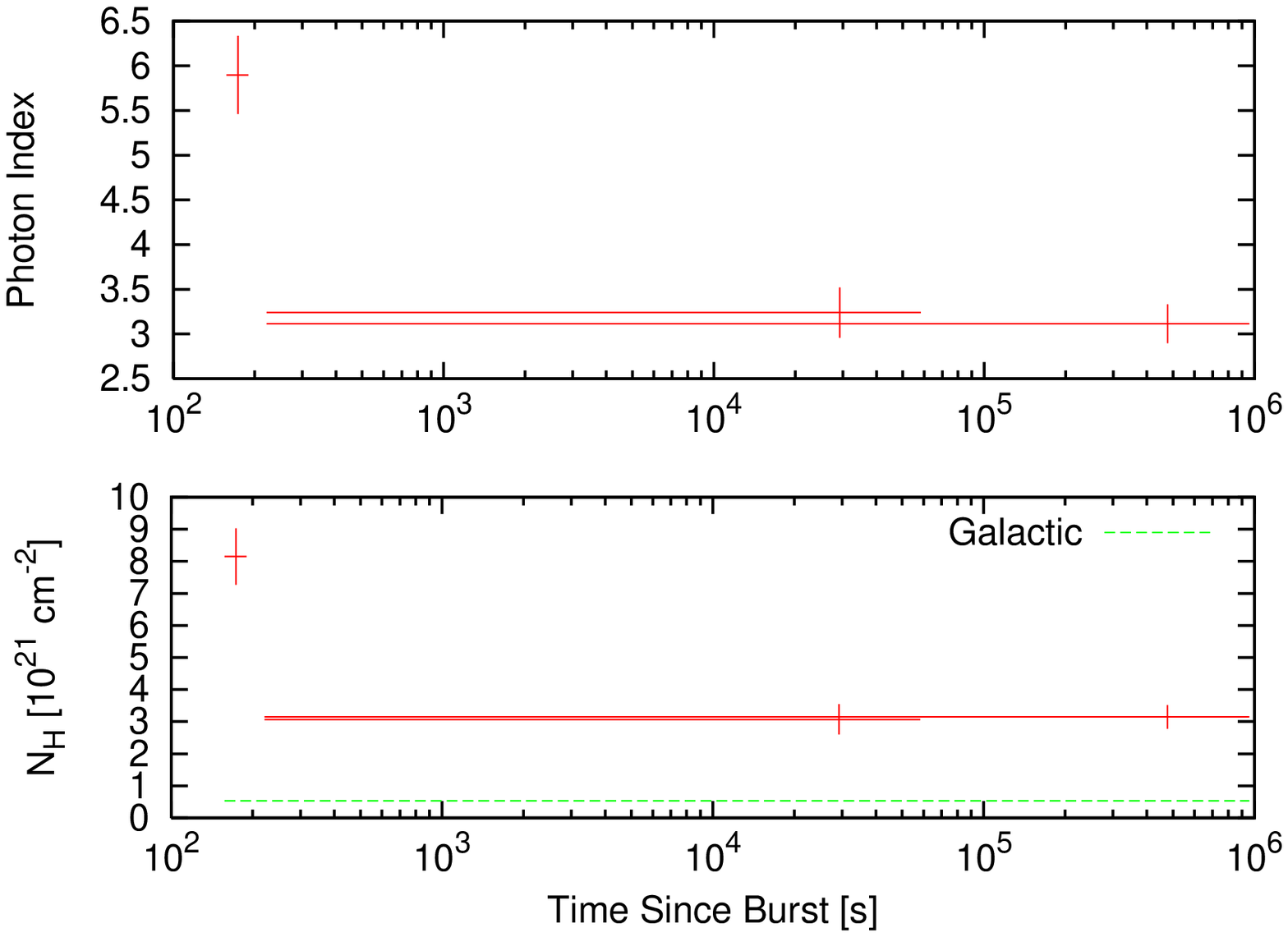}
\caption{\small
(A) The X-ray light curve and hardness plot for GRB~050714B.
(B) There is strong spectral evolution during the early X-ray afterglow.}
\label{fig:050714B_lc}
\end{figure}

Also a clear outlier in Figure \ref{fig:cont}, the WT mode spectrum
for the time interval 157.4 to 189.7s (\#1 in Table 6)
corresponds to a rapid decline in the X-ray light curve
and is well-fit by a blackbody.  Alternatively, the spectrum can be modeled
by a power-law with photon index $\Gamma=5.9\pm 0.4$ \citep[see, also,][]{page05}.
The spectrum during the flare (PC mode; $t\approx 275-525$s) is poorly fit by either a
blackbody ($\chi^2/\nu=97.19/35$) or a power-law ($\chi^2/\nu=76.78/35$).
The PC mode spectrum from the start of the flare until 58.3 ksec, can be marginally well fit
($\chi^2/\nu=48.64/33$) by a power-law plus blackbody.  In Table 6, we also
extract the PC mode data during the flare only (interval \#3) and demonstrate that the same
result holds there, although with larger error bars.  Between regions \#1 and \#3, the
blackbody temperature in the observer frame decreases from 0.2 to 0.1 keV, and the radius increases from $R_{\rm BB}= (2.7\pm0.5) \times 10^{12}$ cm
to $(1.2\pm0.7)\times 10^{13}$ cm (Equation \ref{equation:bbody}), assuming $z=2.66$ (see below).

\begin{figure}[H]
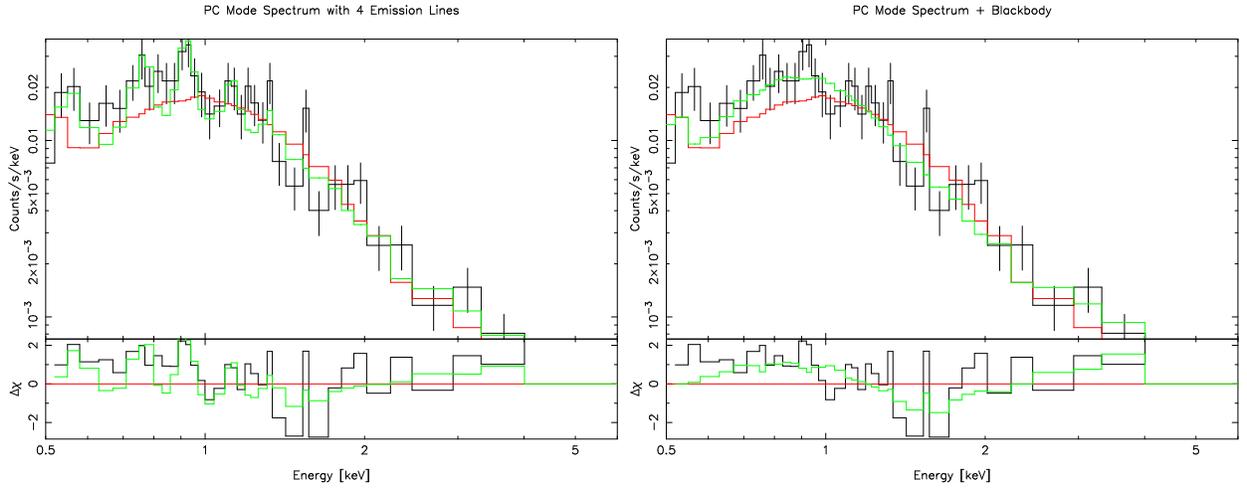

\rotatebox{270}{\includegraphics[width=2.5in]{f12a.ps}}
\rotatebox{270}{\includegraphics[width=2.5in]{f12b.ps}}
\caption{\small
(A) During the GRB~050714B flare at $t\sim 400$s, emission lines at 0.56, 0.76, 0.91, and 1.12 keV (green curve), possibly associated with
Si~XIV, S~XVI, Ar~XVIII, and Ca~XX at $z=2.66$, improve the power-law model fit (red curve) at $4.2\sigma$
significance (Table 7).
(B) The power-law fit is also improved significantly with the inclusion of a blackbody (Table 7).
The residuals panels in each plot show how the additional model components (green curves) fit the power-law continuum model
residuals (black curves).}
\label{fig:050714B_lines}
\end{figure}

Alternatively, the power-law fit to interval \#2 is improved at $4.2\sigma$ significance
with the inclusion of
4 emission lines ($\chi^2/\nu=41.69/27$; $\Delta \chi^2 = 35.09$, $\nu=8$; Figure \ref{fig:050714B_lines}).
The fit to region \#3 is improved less with the addition of lines, however there is a small
number of counts in that hand-selected region, and the line locations are consistent with those
from region \#2 (Table 7).
There are a number of plausible line identifications for region \#2.  For the lines
at 0.91 $\pm$ 0.03, 0.76 $\pm$ 0.05, 0.56 $\pm$ 0.05, and 1.12 $\pm$ 0.05 keV, one association is
made with H-like species Ar~XVIII, S~XVI, Si~XIV, Ca~XX, respectively, at $z=2.66$.
A thermal (MEKAL) plasma with normalization (see, Table 3) $0.1\pm0.05$ for solar abundances and $kT=1.0\pm0.2$ keV
at a similar redshift ($z=2.4$), in addition to the power-law, provides 
an acceptable fit to the data ($\chi^2/\nu=44.48/33$).

Comparing the hypothesis of emission lines to that of a blackbody (in
addition to a non-thermal continuum) for GRB~050714B, we find that this
event is a compelling line emission candidate.
The power-law plus line model yields an excellent fit ($\chi^2/\nu=41.69/27$).
However, this is little improvement over the simpler power-law plus blackbody
model ($\chi^2/\nu=48.64/33$).  The better fit of the MEKAL plasma model ($\chi^2/\nu=44.48/33$) for the
same number of degrees of freedom appears to reinforce the line emission hypothesis.

\subsection{Detections in the Dyadically Grouped, $>500$ Count Spectra}
\label{sec:dyadic}

The sample of spectra under study is roughly doubled when we consider the additional
spectra formed by dyadically grouping the $\sim$ 500 count spectra
into more fully-exposed spectra containing
$>500$ counts.  We observe that the inferred line significance is clearly a function of the
number of counts within the spectrum, with the more fully exposed spectra showing a higher
probability of false trigger.  The median significance (from the LRT) for
1-5 emission lines for the full sample is 1.71$\sigma$.  The lower and upper
quartiles are 1.21$\sigma$ and 2.23$\sigma$, respectively, implying a limit of $5.3\sigma$
for a strong outlier.   We find that only 3 bursts exhibit spectra with line 
triggers at the $>5\sigma$ level: GRBs 060218, 060202, 050822.  The maximal significances
are $20.6\sigma$ ($t=160$s-2.8 ksec; 27,620 cts), $6.5\sigma$ ($t=150$s-1 ksec; 27,620 cts), and
$5.2\sigma$ ($t=482.9-525.5$s; 1050 cts), respectively.  GRBs 060218 and 050822 are 
discussed in detail above.  Here, we briefly present the GRB~060202 spectrum.

\begin{figure}[H]
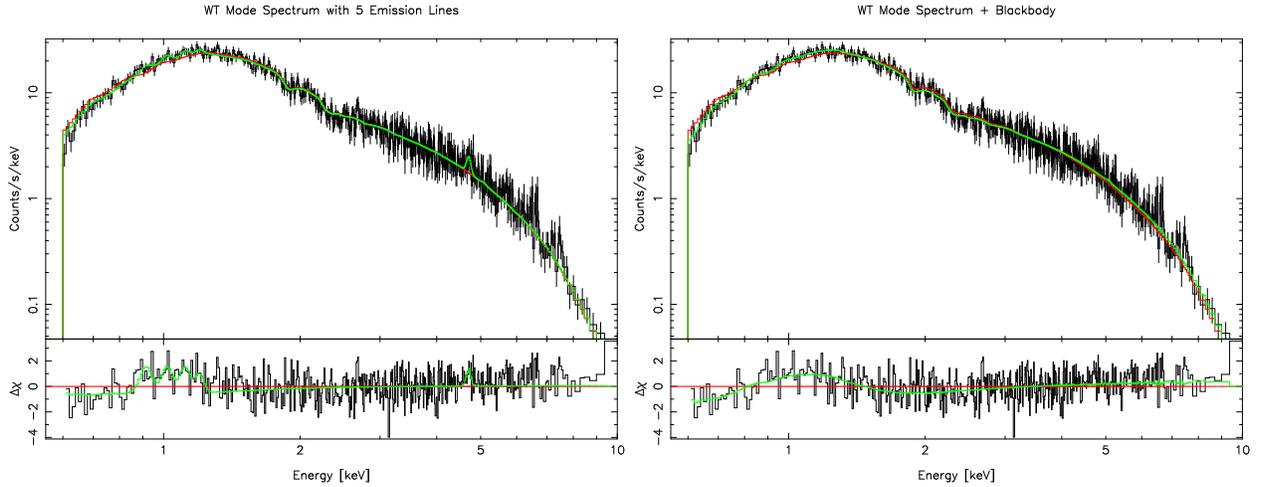

\rotatebox{270}{\includegraphics[width=2.5in]{f14a.ps}}
\rotatebox{270}{\includegraphics[width=2.5in]{f14b.ps}}
\caption{\small
(A) The absorbed power-law model fit (red curve) to the WT mode data 150 to 1000s after GRB~060202 is improved
significantly with the addition of 5 emission lines (green curve) at 0.9, 1.01, 1.12, 1.19, and 4.7 keV, possibly
associated with Al~XIII, Si~XIII, Si~XIV, Ni~XXVIII, and S~XV at $z=0.783$.
(B) A slightly better fit is found by adding a blackbody to the powerlaw.
The residuals panels in each plot show how the additional model components (green curves) fit the power-law continuum model
residuals (black curves).}
\label{fig:060202_lines}
\end{figure}

The WT mode data during the first 1 ksec of GRB~060202 show two temporal declines separated
by a plateau, with mild spectral evolution (Figure \ref{fig:060202_lines}) then and afterward.
The spectrum in the 0.6-10 keV band
from 150s to 1ksec after the burst is unacceptably fit by a blackbody model
($\chi^2/\nu=2511.64/507$).  Less poor is the absorbed power-law model fit ($\chi^2/\nu=644.54/507$), with the
following best-fit parameters:
 $N_H= (5.0 \pm 0.1) \times 10^{21}$ cm$^{-2}$ \citep[significantly in excess of the Galactic column:
$N_{H,{\rm Galactic}} = 5 \times 10^{20}$ cm$^{-2}$;][]{dickey1990},
  $f = (2.04 \pm 0.02) \times 10^{-9}$ erg cm$^{-2}$ s$^{-1}$ [0.5-10 keV], and $\Gamma = 2.32\pm 0.02$.
Consistent values are reported by \citet{morris06}.
The fit is improved at $6.5\sigma$ significance with the inclusion of 5 emission lines
(Figure \ref{fig:060202_lines}; $\Delta \chi^2 = 69.06$, for 10 additional degrees of freedom).  The best fit line
energies are: 0.9 $\pm 0.03$, 1.01 $\pm 0.03$, 1.12 $\pm$ 0.07, 4.7 $\pm 0.07$, and
$(1.19 \pm 0.04)$ keV.  The line equivalent widths are 17 $\pm$ 3, 17 $\pm$ 2, 17 $\pm$ 2,
50 $\pm$ 15, and  10 $\pm$ 2 eV, respectively.  The line luminosities are 1.23 $\pm$ 0.01,
1.4 $\pm$ 0.1, 1.6 $\pm$ 0.1,  0.5 $\pm$ 0.2, and  $(0.9 \pm 0.1) \times 10^{-11}$ erg cm$^{-2}$
s$^{-1}$, respectively.

From Keck optical spectroscopy, we have determined that the host galaxy redshift is $z=0.783$.
The X-ray lines can be indentified with Al~XIII, Si~XIII, Si~XIV, Ni~XXVIII, and S~XV, respectively.
In an initial draft of this paper, we had guessed $z=0.4$ for this event.  Altough the prediction was
not borne out in detail, the complex of X-ray lines near 1 keV did correctly indicate a moderately low $z$.
No lines are detected after 1 ksec at the $>2.5\sigma$ significance
level.  At the 3$\sigma$ level, the line luminosities are below $1.6 \times 10^{-11}$ erg cm$^{-2}$ s$^{-1}$ and the
equivalent widths are below 300 eV.

Alternatively, we can model the soft component by adding a blackbody with $kT=0.18\pm 0.02$ keV (observer frame)
and luminosity
$(7.4 \pm 2.4) \times 10^{-10}$ erg cm$^{-2}$ s$^{-1}$, implying a radius $(3.0\pm0.8) \times 10^{12}$ cm (Equation \ref{equation:bbody}).  
The reduced $\chi^2$ 
is marginally lower than for the power-law plus lines model ($\chi^2/\nu=570.68/505$).  The absorption is
$N_H= (5.9 \pm 0.3) \times 10^{21}$ cm$^{-2}$, and the power-law model parameters are
$f=(1.8\pm 0.1) \times 10^{-9}$ erg cm$^{-2}$ s$^{-1}$ [0.5-10 keV], and $\Gamma=2.20\pm 0.04$.
The thermal component can also be modeled by a MEKAL plasma. However, the implied abundances are zero,
indicating that the data prefer an additional continuum (rather than discrete) component.

\begin{figure}[H]
\includegraphics[width=3.3in]{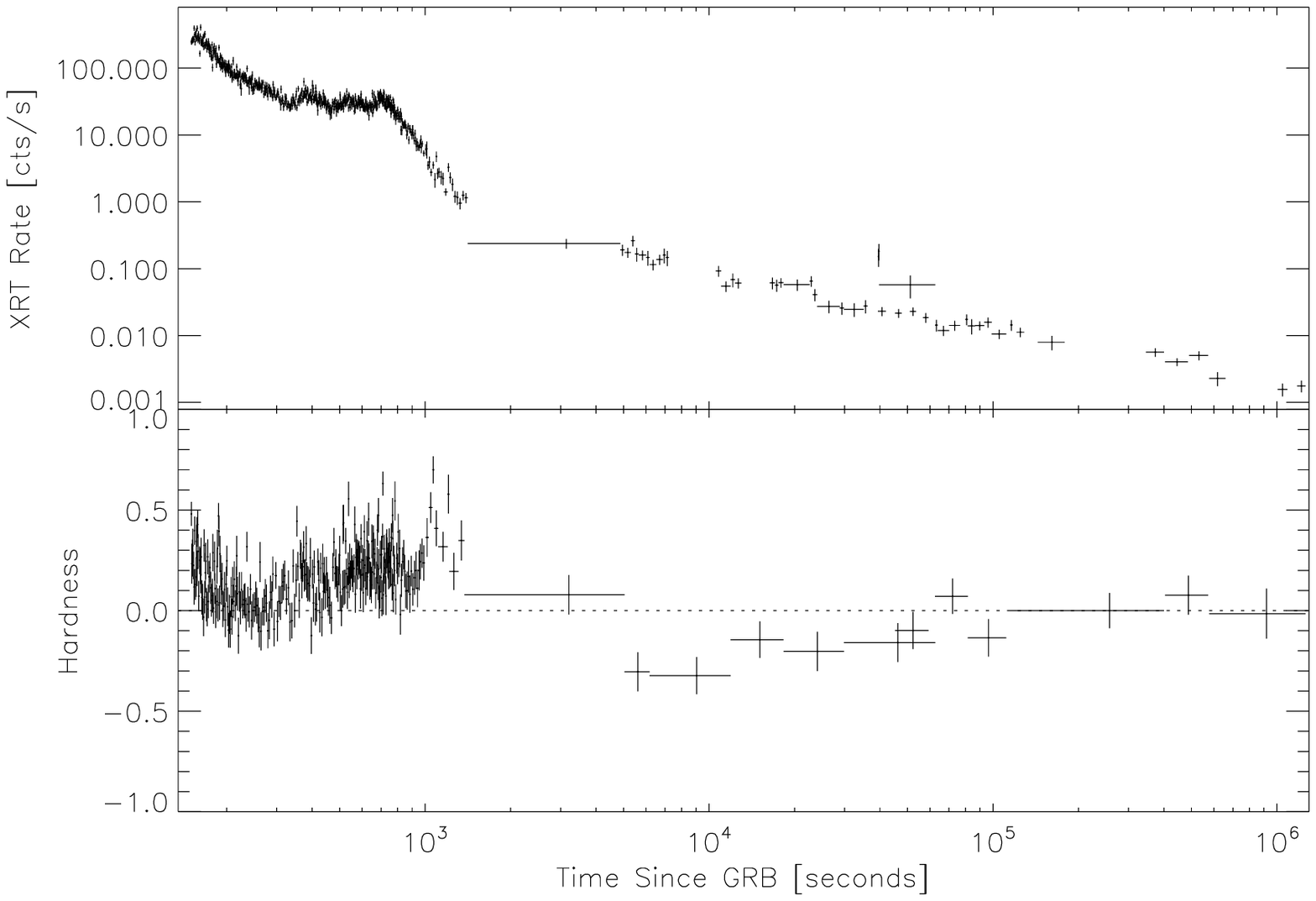}
\includegraphics[width=3.3in]{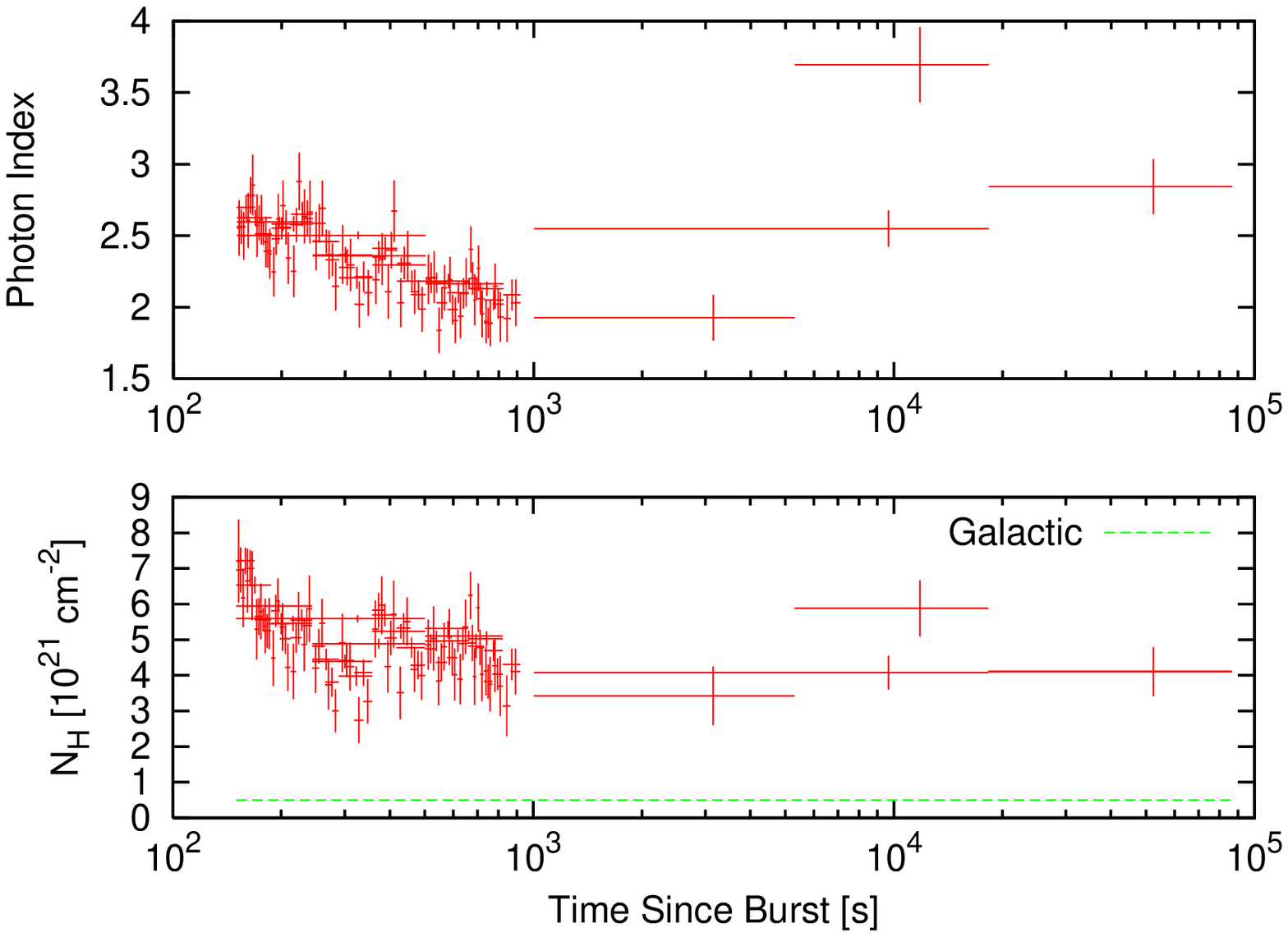}
\caption{\small
(A) The X-ray light curve and hardness plot for GRB~060202. (B) The absorbed power-law model fits.}
\label{fig:060202_lc}
\end{figure}

Comparing the hypothesis of emission lines to that of a blackbody (in
additional to a non-thermal continuum) for GRB~060602, we find that the
blackbody model provides a better fit ($\chi^2/\nu=570.68/505$) than the
more complex power-law plus lines model ($\chi^2/\nu=575.45/497$).

\section{Discussion}

\subsection{The Spectra of Other Bright Flares}
\label{sec:others}

Given the detection of soft X-ray components in two bright flares, it is possible that a correlation
may exist between this emission and the X-ray flaring.
To explore further this possibility, we extract
and fit spectra by hand to the integrated counts of 19 bright (peak count rate $>10$ cts/s) flares (Table 8).
Because a blackbody model provides a poor fit to
each spectrum, we present only the power-law model fits (Table 8).  The $\Gamma$ values for GRBs 050714B and 050228
stand out from the other values, most of which are $\Gamma \sim 2$.  (There is also one hard outlier,
GRB~050820, with $\Gamma = 0.99\pm 0.05$.)  We therefore conclude that these very soft spectra are uncommon in the flares.

A few of the flares, in addition to those discussed above, exhibit marginal significance 
($>3\sigma$) evidence for lines on top of their non-thermal spectra (Table 9): GRBs 050502b, 051117A (during two
flares), 060124 (first flare, with weak detections in the next).
For GRB~060124 at $z=2.296$ \citep{proch06}, there
is a possible Fe~XXVI line near 1.9 keV and several lines at higher energies 
possibly associated with recombination of Fe- group elements.  Although
no redshift is known for GRB~050502B, the lines may also be associated
with the Fe group elements:
 1.44 keV---Fe~XXVI, 1.69 keV---Ni~XXVII, 4.48 keV---?,
 2.14 keV---Fe ionization?  
One possible
association for the lines in GRB~051117A at energies $\sim1.8$, 2.8, 0.6, 1.4, 0.9, and 1.1 keV, is with H- or He-like
Ca, Fe, Ar, Ne, Si, and S, respectively, at $z=1.2$.  We expect $\sim 1$ trigger at $\gtrsim 3\sigma$ in 22 trials.  The actual number of
triggers (6) suggests a modest increase in the line emission probability during the bright flaring episodes relative to the quiescent periods
(e.g., Figure \ref{fig:multi_probs}).

\subsection{Blackbody Emission: Shock Breakout Through the Progenitor Star?}
\label{sec:bbody}

The optical/IR emission from GRB~060218 has yielded unequivocal evidence for an underlying type-Ic supernova--SN~2006aj \citep{mirabal06,soller06,modjaz06}.
It has been suggested recently \citep{campana06} that the unusual early X-ray emission is due to the propagation of a radiation dominated
shock through a wind or H envelope surrounding the progenitor star.
In this picture, the early X-ray light curve is broad, then rapidly declining, due
to light travel time effects across an aspherical shock shell of thickness $\sim R/\Gamma c \sim 300$s.

The shock can be due to the GRB itself \citep{tan01,colgate74} or due to mildly relativistic material 
expected to form and surround the GRB jet as it punches through
the He core of the progenitor star at $R_{\rm He}\lessim 10^{11}$ cm \citep{mnr01,rcr02}.
After the shock punches through an optically thick layer around the He core and interacts with the envelope, it leaves behind an expanding photosphere.
The radius at which the light is able to escape the photosphere around a carbon-rich, Wolf-Rayet star
is $R_p = 1.5\times 10^{12} { \dot M_{-4} \over v_{w,8} }$ cm \citep{tan01}.
The scatter in observed mass loss rates 
($\dot M_{-4}$ in units of $10^{-4} M_{\odot}$ yr$^{-1}$) and wind velocities ($v_{w,8}$ in units of $10^8$ cm/s) is $\sim 1$ dex 
\citep{koest95} and could lead to a $\sim \pm 1$ dex range in $R_p$.
Due to the energy input ($\sim 10^{51}$ erg) the expected photospheric blackbody temperature is $\sim 0.3$ keV.

In the course of our line search, we find spectra from 3 other events which are soft and well fit by models 
containing a blackbody.  Aside from no clear detection in the UV \citep{gronwall05,page05b,blustin06}, these detections are similar
to GRB~060218 in a number of ways.
In each case the observation epoch is $\sim 1$ ksec, and the blackbody radii are $R_{\rm BB} \sim 10^{12}-10^{13}$ cm.
We also find similar temperatures ($\sim 0.1-0.5$ keV) in each case, which appear to decrease in time mildly.
In all cases, the emission is transient.
The $R_{\rm BB}/c$  are similar to the light travel time from the GRB to the observation epoch, which would imply a simultaneous GRB and SN
in either scenario discussed above.
The radii are at least 2 orders of magnitude smaller than the distance expected for the highly relativistic ($\Gamma\sim 10$) external shock 
at the observation epoch, ruling out an association with the external shock.

\begin{figure}[H]
\centerline{\includegraphics[width=5.0in]{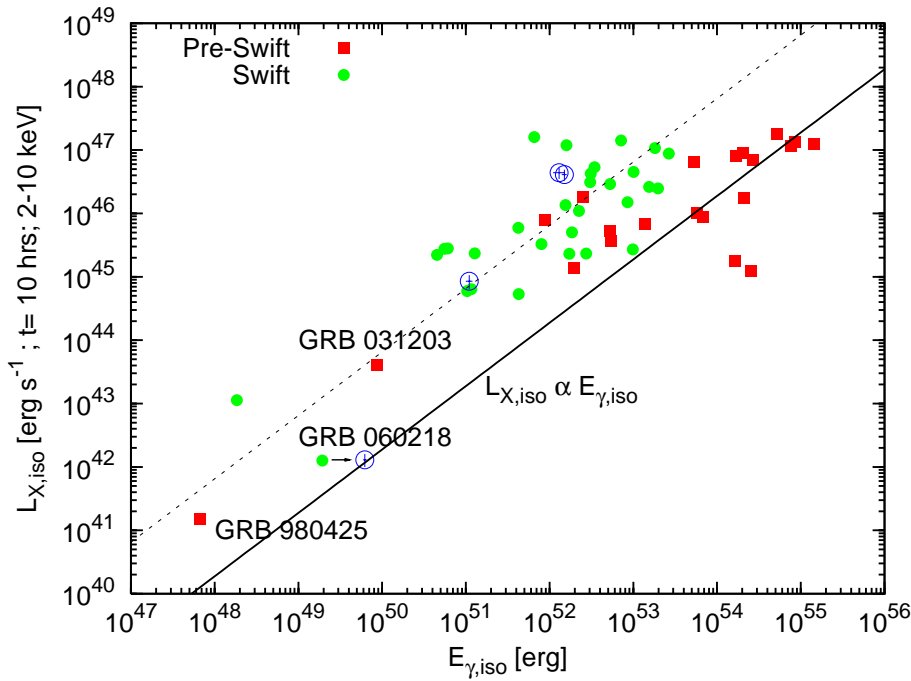}}
\caption{\small
The isotropic X-ray luminosity at $t=10$ hours in the host frame versus the isotropic prompt energy
release in $\gamma-$ray's.  Fluence data for {\it Swift}~(typically in the 15-150 keV band) are taken from the table at
http://swift.gsfc.nasa.gov/docs/swift/archive/grb\_table, and these are translated to $E_{\gamma,{\rm iso}}$
by multiplying by $4\pi D_L^2/(1+z)$ only.  The bursts studied here are shown encircled
in blue.  The parallel track
set by the {\it Swift}~bursts relative to bursts prior to {\it Swift}~likely reflects only the narrow BAT band
pass and our non-application of a k-correction, which apparently leads to systematic
underestimates of $E_{\gamma,{\rm iso}}$.
Note the shift in $E_{\gamma,{\rm iso}}$ for GRB~060218 depending on whether we use the
tabulated value or the more carefully calculated value from \citet{campana06}.}
\label{fig:lx}
\end{figure}

An additional clue to the mechanism which produces the soft component may come from a simple comparison of the GRB and shock kinetic energies.
We can infer the shock kinetic energy using the X-ray afterglow luminosity at $t=10$ hours in the host frame, which is expected to scale linearly
with the shock energy due to an independence on the density and a mild dependence on the shock microphysical parameters \citep[e.g.,][]{bkf}.
GRB~060218, like GRB~980425/SN~1998bw and GRB~031203/SN~2003lw, is found to track other GRBs from {\it Swift}~and previous missions
(Figure \ref{fig:lx}),
implying a roughly similar efficiency for the conversion of shock kinetic energy into $\gamma-$ray's, but with a substantial sub-luminosity 
evident in both quantities.  The other bursts discussed above also follow the trend $L_{\rm X,iso} \propto E_{\gamma,{\rm iso}}$.  

On the other hand,
using the fact that the duration of the photospheric expansion phase appears to be $\sim 10^2-10^3$s for each burst, we can estimate the ratio of
blackbody fluence to prompt GRB fluence (Table 10).  The prompt $\gamma-$ray energy
release for GRBs 060202, 050822, and 050714B is $S_{\gamma}=2.4\pm 0.2$ \citep{hull06}, $3.4 \pm 0.3$ \citep{hull05}, and 
$(0.7\pm 0.1) \times 10^{-6}$ erg cm$^{-2}$ \citep{tuell05}, respectively, in the 15-150 keV band.  
The ratios are found to be: $0.3\pm0.1$, $0.03\pm0.01$, and $0.2\pm0.1$, respectively.  
The ratio value for GRB~060218 is $4\pm 1$.   The soft fluence clearly correlates with the prompt $\gamma-$ray and shock emission.
The large dispersion over more than two orders of magnitude could be explained by the
diversity of observed progenitor wind outflows and the strong dependence of the available energy $E$ on the mass loss history:
$E \approx \Gamma^2 R_p { \dot M c^2 \over v_w } \propto ({\dot M \over v_w })^2$ \citep{tan01}.  

Because wide variations in the thermal energy release apparently do not alter the  $L_{\rm X,iso} \propto E_{\gamma,{\rm iso}}$
relation, we conclude that the soft component is more likely
to have been produced by the cocoon surrounding the GRB jet as it escape the progenitor He core than by the GRB jet itself.
A radiation dominated cocoon is not expected to appreciably affect the standard internal shocks or afterglow emission \citep{rcr02}.
On the issue of the faintness of GRB~060218, \cite{fan06} discuss the possibility that GRB~060218 was an almost ``failed GRB,'' 
with a large fraction of the GRB energy not making it out of the
progenitor star and winding up as thermal photons in the cocoon.

One additional insight which comes from the extreme proximity of GRB~060218, it's sub-luminous $\gamma-$ray emission, and the relative faintness of 
the soft X-ray components in 3 other bursts
is that this emission could be relatively common.  Bursts like GRB~060218, where the spectrum is dominated by the soft component could outnumber normal 
GRBs 100 to 1.  It may be a selection effect that the blackbody radiation happens to peak where the detector response is also maximal at $E\approx 5kT\sim 0.5-2$ keV
for each of these events; soft or harder spectra due to breakout at larger or smaller radii, respectively, may also occur.

\subsection{Constraints on the Line Emission Mechanisms}
\label{sec:mech}

In the previous section we discussed the soft X-ray emission in terms of the blackbody fits.  For 3 of the 4 bursts in Section \ref{sec:bigthree}, the 
data appear to be fit better with an emission line model.  Clear line associations are possible for GRB~060218 and possible
association are presented for the other bursts lacking redshift measurement.  For two of the bursts, the line component is
well fit by a thermal plasma model containing only two parameters (temperature and normalization).  Because, the continuum under the
lines is dominated by a power-law, the emitting plasmas may well not be in thermal equilibrium.  \citet{laz03} has discussed
the relative unlikelihood that GRB X-ray lines could be due to thermal plasmas.  For simplicity in fitting, however, the MEKAL
code generates approximately correct line locations and sensible flux ratios for a broad range of astrophysically abundant elements.
The code has been shown to produce similar lines as from photo-ionization models \citep[e.g.,][]{watson03}, at the level
of spectral resolution appropriate to {\it Swift}~XRT data.  In the case of GRB~050822, the MEKAL model cannot produce the closely spaced lines,
and this may argue for a non-equilibrium ionization state, for two plasmas with differing ionization states, or for
plasmas moving at different speeds ($\lessim 0.1c$) relative to the observer.

In none of the cases do we detect lone, statistically significant emission lines.  Rather, we find sets of lines, typically fairly closely
spaced in energy, which we associate with ionized light metals or with L-shell transitions in Fe (GRB~060218).  
The line sets we find are perhaps most similar to the light metal lines \citep{reeves02,watson03} and soft, possibly thermal excesses
\citep{watson02} found in {\it XMM}~data.  A detection of such lines has been claimed for {\it Chandra}~as well \citep{butler03}.

\begin{figure}[H]
\centerline{\includegraphics[width=5.0in]{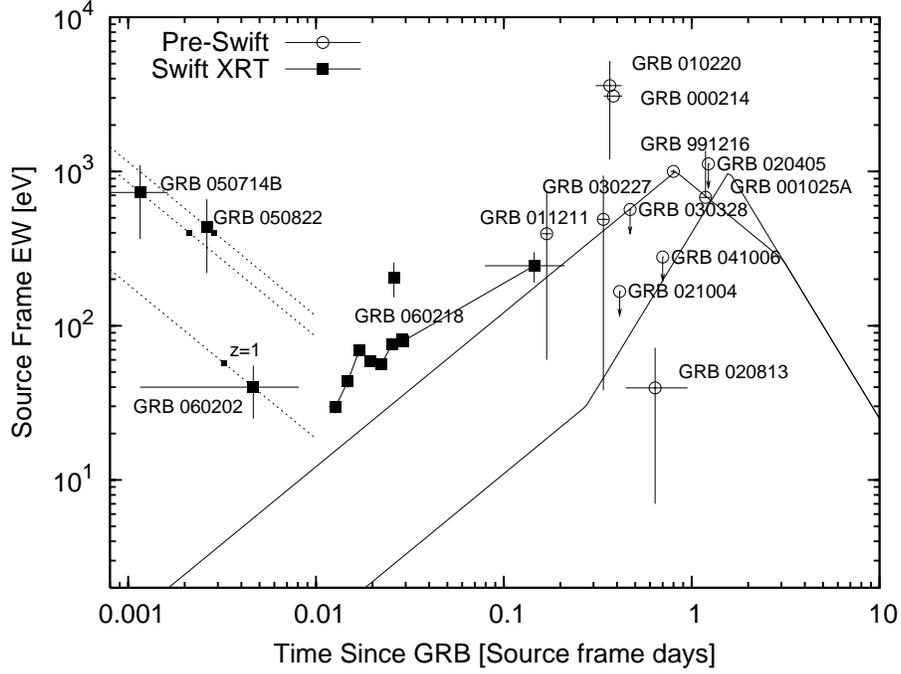}}
\caption{\small
Equivalent width $EW$ versus time in the source frame, adapted from \citet{butler05b}.  The four bursts studied here are marked as solid squares.
The dotted curves show how the point for GRBs 060202, 050822, and 050714B move as the redshift is varied (small solid squares mark $z=1$
for each burst).  A number of points are measured for GRB~060218 in the $\sim $16,000 cts spectra (connected by a line), with one possible
larger $EW$ value from the $\sim 500$ cts spectra (Section \ref{sec:bigthree}).  Previous detections and upper
limits
are marked with open circles and labeled.  The solid curves represent photoionization models from \citet{bnr01}, explained in more detail
in \citet{butler05b}.}
\label{fig:eqwid}
\end{figure}

\begin{figure}[H]
\centerline{\includegraphics[width=4.0in]{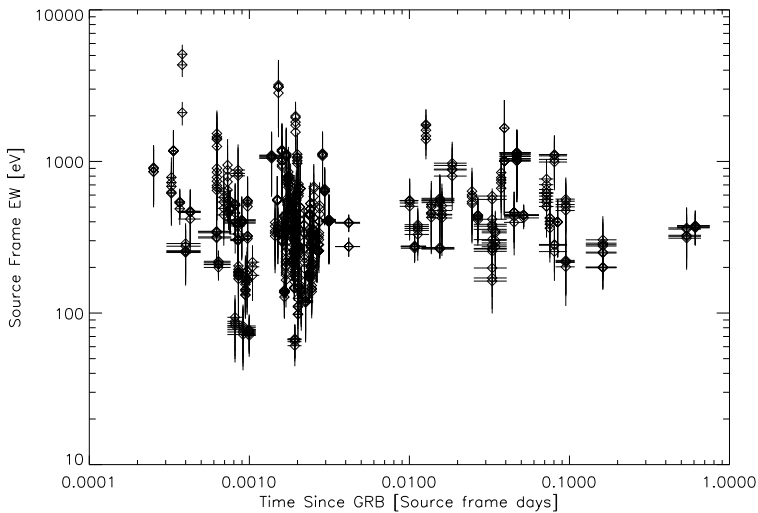}}
\caption{\small
Equivalent width $EW$ versus time in the source frame for all (marginal) triggers at $>2\sigma$ significance for 28 {\it Swift}~XRT afterglows with measured
redshifts, excluding GRB~060218.  Redshifts are taken from the GCN.  We restrict to only the spectra observed with exposures $\delta t< t_{\rm mean}/2$.
Ninety percent of the $EW$ values below 0.1, 0.01, or 0.01 days have $EW<1$ keV.
There are few spectra (3\% of the sample) at times $>0.1$ day in the source frame.}
\label{fig:ew}
\end{figure}

Figure \ref{fig:eqwid}, adapted from \citet{butler05b}, displays the line triggers in the context of lines claimed from previous
missions.  For comparison and for estimation of upper limits for the $\sim 90$\% of spectra not showing significant evidence for lines,
Figure \ref{fig:ew} displays equivalent width values versus time for marginal detections in the full dataset of {\it Swift}~bursts with measured redshift.
It is difficult to use the triggers or upper limits to rule on the reality of the historic lines, because there are few well-exposed XRT spectra at $t\gtrsim 0.1$ day.
However, observations with {\it Chandra}~place stringent limits on the late-time lines \citep[down arrows in Figure \ref{fig:eqwid};][]{butler05b}.
The upper limits from the non-detections ($EW\lessim 1$ keV at $t<0.1$ day) in the {\it Swift}~XRT sample are roughly consistent with the detections.

The most striking feature of the triggers is that the possible lines are emitted at dramatically earlier times ($t<0.1$ day) than in previous cases.
Prior to {\it Swift}, it was suspected that GRB lines would be very difficult to form at early times
\citep[aside from exotic possibilities suggested by, e.g.,][]{gmk05} due to the overwhelming flux from the bright X-ray afterglow.  The solid
lines in Figure \ref{fig:eqwid} show the rapid decrease expected at early times for two photo-ionization models \citep{bnr01,butler05b}.
The data from GRB~060218 may roughly follow this expectation, but the other data likely do not.

One of the big surprises from {\it Swift}~was the departure of the early X-ray afterglow light curves from late time extrapolations
\citep[early flaring, anomalously flat and rapidly decaying early light curves, e.g.,][]{nousek06,zhang05}.  In some cases, the afterglows
are fainter at early times ($t\sim 10^2 -10^4$s) than previously suspected.  
The explanation for why we see lines possibly correlated with flare
emission may simply be that both depend on similar circumstances---non overpowering external shock emission---for their detection.

Both the trigger times and apparent transient nature of the lines limit the distance from the progenitor:
\begin{equation}
\label{equation:radius}
R < {ct_{\rm obs} \over 1+z} {1 \over 1-\cos(\theta) } \lessim 10^{13} {\rm ~cm,}
\end{equation}
where $\theta$ is the angle between the line emitting material and the line of sight \citep[e.g.,][]{vietrietal01}.  This implicates models from the
so called ``nearby reprocessor'' class, whereby the lines are produced by a long-lasting central engine \citep{rnm00} or with the help 
of magnetic field energy stored in the plasma bubble discussed above \citep{mnr01}.  The small distances from the progenitor
favor one-step ``hypernova'' explosions
\citep[e.g.,][]{woosley93,pacz98} and disfavor a SN occurring prior to the GRB \citep[e.g,][]{vietri98}.  For the plasma bubble,
a possibly large magnetic field ($\sim 10^5$ Gauss) can be entrained and carried with the flow, both maintaining clumps of matter
in the flow and allowing for synchrotron irradiation of the clumps \citep{mnr01}.  Or the lines could be generated by reflection from cold
matter in the funnel wall carved out by the GRB jet from the progenitor \citep[e.g.,][]{kmr03}.

The chemical species we infer point to ionization parameters $\xi = L_{\rm x}/nR^2 \lessim 10^2$ \citep{lrr02}.
If the continuum luminosities $L_{\rm X}$ we measure are directly responsible for the lines, this implies a density $n\sim 10^{18}$  
cm$^{-3}$ (GRB~060218) - $10^{21}$ cm$^{-3}$ (GRB~050822).  Because the largest densities expected in the nearby reprocessor scenario are
$\gtrsim 10^{17}$ cm$^{-3}$ \citep{rnm00,mnr01}, a small fraction of the observed continuum ($\sim 0.01-10$\%) must be powering the lines.

\subsection{Systematic Uncertainties in the XRT Response}
\label{sec:cal}

Calibration efforts for the XRT spectral response are ongoing\footnote{http://swift.gsfc.nasa.gov/docs/swift/analysis/xrt\_digest.html}.  
The quality of the current response matrices
at low energy could have an important impact on the explanation for the soft excesses, particularly
on the possibility of X-ray lines.  Here we address a number of concerns raised by the XRT team (D. Burrows, private communication).
This can be done quantitatively now due to the release of the v008 response files.  The v008 files allow for the fitting of both the
PC and WT mode data in the full 0.3-10 keV band, with the requirement of a small (3\%) systematic error component.

We have rerun the analysis using the v008 response files and including the systematic errors.  For the 0.3-10 keV band, we find consistent 
distributions for
the continuum fit parameters and continuum fit goodnesses (Figure \ref{fig:cont}) 
and line trigger significances (Figure \ref{fig:multi_probs}) for the PC mode data.  The distributions
for continuum fit parameters for the WT mode data are also consistent.  There
is, however, a dramatic increase in the number of line triggers for the WT mode data for the well exposed spectra of 5 bright afterglows
(GRBs 050502B, 060124, 060202, 060210, 060211A).  The line triggers fall near 0.5 keV and are clearly due to a dip near the O-K edge at 0.532 keV, 
which can also been see in Figures \ref{fig:050714B_lines} and \ref{fig:060218_whole} for the spectra for GRBs 050714B and 060218.  
The XRT team argues that the dip---also seen in calibration observations of the Crab, 3C273, and H1426$+$428---is due to a gain offset $\lessim 80$ eV, 
which occurs occasionally for WT mode data due to a problem with the bias subtraction.  (It may also occur for the PC mode data, possibly due to
illumination by the bright Earth.)  Because we observe the magnitude of the dip to be proportional to the absorbing column $N_H$ (hence
to the depth of the O-K edge), we agree with this explanation.  This observation rules out the more pernicious possibility of a significant problem 
with the response files below 0.6 keV.  Also, we note that the spectra for the 5 WT mode bursts showing the dip can be adequately fit by absorbed
power-laws if we also fit for the gain offset.

One possible additional concern with 3 of the 4 bursts with sigificant line triggers is that the triggers appear to occur near the same observed
energies (i.e., at $\sim $100 eV intervals near 0.8, 0.9, 1.0 keV, etc), possibly indicating an instrumental origin.  From simulations,
we note that line-like residuals due to the gain offset are unimportant $\lessim $10\% effects in this energy range.  (There can be narrow residuals
near the Si-K edge at 1.839 keV at the $\sim$ 20\% level, however).  Also, fitting for the gain, we find that the inferred trigger 
significances do not change.  We, therefore, rule out the possibility that the instrumental effect of the gain offset produces the line triggers.
Reinforcing this conclusion, a large number spectra with comparable source intensities to those with possible line detection are well
fit by simple power-laws (Figures \ref{fig:cont}A, \ref{fig:multi_probs}).
However, we cannot entirely rule out the possibility that the low spectral resolution of the detector could allow for acceptable fits 
of an additive continuum component by closely spaced lines.  This possibility is strengthened by the degradation of the
of spectral resolution (from 0.08 keV at launch to $\sim$ 0.1 keV in late 2006) due to radiation damage, an effect which is not treated
in the v007 or v008 response files.

\section{Conclusions}

We have conducted a thorough and blind search for emission lines through nearly 0.4 years of X-ray afterglow
data accumulated by the {\it Swift}~XRT.  The majority of spectra (90\%) are well fit by absorbed power-law models, but significant
outliers to the population at the 5-10\% level exist and have anomalously soft, possibly thermal spectra.  Four bursts are singled out as possibly
exhibiting
2-5 emission lines: GRBs 060218, 060202, 050822, and 050714B.  Removed from the sample, line triggers in the spectra of the other 
bursts are consistent with Poisson fluctuations in the continua.

The most significant
soft component detections in the full data set of $\sim 2000$ spectra correspond to GRB~060218/SN~2006aj, with line triggers ranging
from $4\sigma$ to $\sim 20\sigma$.  A thermal plasma model fit to the data indicates that the emission may be primarily due to L-shell transitions of
Fe at $\sim$ solar abundance.  The possible line emission occurs near $t\sim 1$ ksec, with a similar observed duration, indicating emitting
material at $R\sim 10^{13}$ cm.  We suggest that the emission is due to the mildly relativistic cocoon of matter surrounding the GRB
jet as it penetrates and exits the surface of the progenitor star.  We associate the ($>4\sigma$ significant) line emission from 
3 other events lacking redshifts with K-shell transitions in light metals.  The lines in these bursts point to emission at similar 
distances, possibly at similar densities $\sim 10^{17}$ cm$^{-3}$, and possibly subject to similar fluxes of ionizing radiation.

As an alternate possibility---difficult to distinguish with the broad XRT spectral resolution ($\sim 80$ eV FWHM at 1 keV)---we 
successfully model the spectra using blackbody continuum components in addition to power-laws.   With radii
(again $\sim 10^{12}-10^{13}$ cm) and temperatures ($\sim 0.1-0.5$ keV) implied by the fits, the possibility exists
that the emission is continuous rather than discrete and possibly again due break out of the GRB shock or plasma
cocoon from the progenitor star.
We find that the energetics of the GRB and its shock, inferred from the 4 events where breakout emission may be present,
point toward the cocoon as the likely source of the soft component.  
Bursts faint in $\gamma-$rays with spectra dominated by a soft X-ray flux possibly due to the shock
breakout may outnumber classical GRBs 100-1.  Typically, the breakout flux would be faint or dominated by the external shock afterglow emission.
The degeneracy between continuum and discrete emission components could possibly be lifted if redshifts are determined for the events discussed
above or if redshifts are measured for bursts detected by {\it Swift}~in the coming years with soft X-ray anomalies.

\acknowledgments
N. Butler gratefully acknowledges support from a Townes Fellowship at U.~C. Berkeley Space Sciences Laboratory and partial support
from J. Bloom and A. Filippenko.  Special thanks to J. Bloom and the U.~C. Berkeley GRB team for comments on the manuscript and several useful
conversations.  Additional thanks to the {\it Swift}~team for impressively rapid public release and analysis of the XRT data.

\input tab1.tex
\input tab2.tex
\input tab3.tex
\input tab4.tex
\input tab5.tex
\input tab6.tex
\input tab7.tex
\input tab8.tex
\input tab9.tex
\input tab10.tex

\end{document}

%% file: tab1.tex
\begin{table}[H]
\begin{center}
\caption{Time-resolved Spectroscopy in $\sim 16,000$ cts Spectra for GRB~060218}
\vspace{5mm}
\tiny
\begin{tabular}{ccccccccc}\hline\hline
\# &  Time (ksec) & BB Flux & $kT$ (keV) & $\chi^2/\nu$ & PL Flux & $\Gamma$ & $N_H$ & $\chi^2/\nu$ \\\hline
   & WT Mode & & & & & & & \\\hline
 1 & 0.161-0.475 & 211.00 $\pm$ 1.72 & 0.94 $\pm$ 0.01 & 1718.48/535 & 372.40 $\pm$ 3.61 & 1.54 $\pm$ 0.02 & 3.97 $\pm$ 0.10 & 595.99/535 \\
   &             & 124.30 $\pm$ 53.90 & 0.18 & $+$ & 324.80 $\pm$ 7.37 & 1.50 $\pm$ 0.04 & 5.73 $\pm$ 0.54 & 537.20/530 \\
 2 & 0.475-0.688 & 40.28 $\pm$ 2.71 & 0.98 $\pm$ 0.01 & 1662.32/549 & 571.20 $\pm$ 5.47 & 1.46 $\pm$ 0.02 & 3.72 $\pm$ 0.10 & 573.38/549 \\
   &             & 147.18 $\pm$ 70.86 & 0.17 & $+$ & 496.20 $\pm$ 10.19 & 1.44 $\pm$ 0.04 & 5.31 $\pm$ 0.54 & 535.00/544 \\
 3 & 0.688-0.872 & 45.30 $\pm$ 3.06 & 0.95 $\pm$ 0.01 & 1724.45/529 & 661.90 $\pm$ 6.35 & 1.57 $\pm$ 0.02 & 4.11 $\pm$ 0.10 & 629.89/529 \\
   &             & 340.95 $\pm$ 152.11 & 0.16 & $+$ & 591.50 $\pm$ 13.57 & 1.55 $\pm$ 0.04 & 6.21 $\pm$ 0.56 & 537.88/524 \\
 4 & 0.872-1.046 & 48.00 $\pm$ 3.22 & 0.92 $\pm$ 0.01 & 1499.49/522 & 689.20 $\pm$ 6.76 & 1.62 $\pm$ 0.02 & 3.97 $\pm$ 0.10 & 640.10/522 \\
   &              & 761.00 $\pm$ 348.78 & 0.13 & $+$ & 659.20 $\pm$ 16.94 & 1.71 $\pm$ 0.04 & 6.85 $\pm$ 0.59 & 603.46/517 \\
 5 & 1.046-1.223 & 44.07 $\pm$ 2.96 & 0.87 $\pm$ 0.01 & 1444.86/500 & 670.20 $\pm$ 6.87 & 1.80 $\pm$ 0.02 & 4.49 $\pm$ 0.10 & 621.45/500 \\
   &              & 2067.00 $\pm$ 851.39 & 0.12 & $+$ & 691.80 $\pm$ 20.88 & 1.94 $\pm$ 0.04 & 8.26 $\pm$ 0.56 & 539.78/495 \\
 6 & 1.223-1.411 & 39.48 $\pm$ 2.65 & 0.82 $\pm$ 0.01 & 1486.17/493 & 622.80 $\pm$ 6.66 & 1.91 $\pm$ 0.02 & 4.46 $\pm$ 0.10 & 602.29/493 \\
   &               & 1992.59 $\pm$ 849.83 & 0.12 & $+$ & 652.90 $\pm$ 21.46 & 2.03 $\pm$ 0.04 & 8.04 $\pm$ 0.57 & 529.29/488 \\
 7 & 1.411-1.617 & 33.38 $\pm$ 2.23 & 0.77 $\pm$ 0.01 & 1419.64/470 & 545.10 $\pm$ 6.16 & 2.03 $\pm$ 0.02 & 4.47 $\pm$ 0.09 & 584.09/470 \\
   &               & 6199.44 $\pm$ 2325.48 & 0.10 & $+$ & 640.30 $\pm$ 22.94 & 2.25 $\pm$ 0.04 & 9.24 $\pm$ 0.50 & 458.65/465 \\
 8 & 1.617-1.851 & 27.12 $\pm$ 1.83 & 0.71 $\pm$ 0.00 & 1413.91/440 & 478.90 $\pm$ 6.18 & 2.23 $\pm$ 0.02 & 4.64 $\pm$ 0.09 & 566.43/440 \\
   &               & 4650.94 $\pm$ 1842.21 & 0.11 & $+$ & 568.40 $\pm$ 23.49 & 2.42 $\pm$ 0.04 & 8.83 $\pm$ 0.52 & 446.16/435 \\
 9 & 1.851-2.116 & 22.03 $\pm$ 1.48 & 0.64 $\pm$ 0.00 & 1563.68/422 & 406.90 $\pm$ 5.56 & 2.34 $\pm$ 0.02 & 4.33 $\pm$ 0.09 & 578.89/422 \\
   &               & 4248.35 $\pm$ 1666.76 & 0.11 & $+$ & 483.30 $\pm$ 21.47 & 2.50 $\pm$ 0.04 & 8.59 $\pm$ 0.51 & 442.47/417 \\
10 & 2.116-2.401 & 18.26 $\pm$ 1.30 & 0.57 $\pm$ 0.00 & 1845.22/389 & 380.70 $\pm$ 6.16 & 2.57 $\pm$ 0.03 & 4.45 $\pm$ 0.09 & 640.99/389 \\
   &               & 3146.32 $\pm$ 1199.94 & 0.11 & $+$ & 417.70 $\pm$ 20.08 & 2.61 $\pm$ 0.05 & 8.10 $\pm$ 0.48 & 400.39/383 \\
11 & 2.401-2.753 & 14.43 $\pm$ 0.98 & 0.53 $\pm$ 0.00 & 1791.20/371 & 313.00 $\pm$ 5.57 & 2.69 $\pm$ 0.03 & 4.34 $\pm$ 0.08 & 630.66/371 \\
   &               & 2706.92 $\pm$ 986.90 & 0.11 & $+$ & 337.90 $\pm$ 16.90 & 2.71 $\pm$ 0.05 & 7.88 $\pm$ 0.45 & 344.64/365 \\\hline
 & PC Mode & & & & & & & \\\hline
12 & 5.950-2.872 Msec & 0.009 $\pm$ 0.001 & 0.36 $\pm$ 0.02 & 215.47/93 & 0.031 $\pm$ 0.003 & 3.41 $\pm$ 0.13 & 4.39 $\pm$ 0.27 & 144.03/93 \\
   &                    & 0.43 $\pm$ 0.16 & 0.10 $\pm$ 0.01 & $+$ & 0.03 $\pm$ 0.01 & 3.17 $\pm$ 0.24 & 7.33 $\pm$ 1.36 & 87.94/91 \\\hline
\end{tabular}
\end{center}
{\footnotesize Notes: Column density $N_H$ measured in units of $10^{21}$ cm$^{-2}$.  $N_{H,\rm Galactic}=1.11 
\times 10^{21}$ cm$^{-2}$ \citep{dickey1990}.  $N_H$ fixed at lower limit $3.7\times 10^{20}$ cm$^{-2}$ for pure blackbody (BB) fits.
Flux measured in units of $10^{-11}$ erg cm$^{-2}$ s$^{-1}$, 0.5-10 keV
for the power-law (PL) and bolometric for the blackbody.  Temperature measured in the source frame, $z=0.033$ \citep{mirabal06}.}
\label{tab:060218s}
\end{table}

%% file: tab2.tex
\begin{table}[H]
\begin{center}
\caption{Time-resolved Line Search in $\sim 16,000$ cts Spectra for GRB~060218}
\vspace{5mm}
\footnotesize
\begin{tabular}{ccccc}\hline\hline
 \# &  Time  & N  & Signif. & Line Energy , Equivalent Width , Flux \\
     &  [s]     &  lines      &     &  (keV,eV,$10^{-11}$ erg cm$^{-2}$ s$^{-1}$) \\\hline
 & WT Mode & & & \\\hline
 1 & 0.161-0.475 & 0 & $<5.0\sigma$ & ... \\
 2 & 0.475-0.688 & 0 & $<5.0\sigma$ & ... \\
 3 & 0.688-0.872 & 0 & $<5.0\sigma$ & ... \\
 4 & 0.872-1.046 & 0 & $<5.0\sigma$ & ... \\
 5 & 1.046-1.224 & 5 & 7.0$\sigma$ & (0.93,28,5.5) (3.19,45,3.3) (3.65,40,2.6) (1.02,20,3.6) (2.76,30,2.5)  \\
 6 & 1.224-1.411 & 5 & 5.9$\sigma$ & (0.95,26,5.2) (1.14,16,2.8) (3.24,28,1.9) (0.87,23,4.8) (1.03,16,3.0)  \\
 7 & 1.411-1.618 & 5 & 8.2$\sigma$ & (0.89,42,9.1) (0.80,40,9.7) (1.06,22,3.9) (4.04,47,2.1) (0.97,18,3.5)  \\
 8 & 1.618-1.851 & 5 & 8.0$\sigma$ & (1.00,37,7.3) (0.79,51,13.4) (0.90,35,7.9) (1.12,23,3.9) (0.71,47,14.1)  \\
 9 & 1.851-2.116 & 5 & 9.3$\sigma$ & (0.96,34,6.0) (0.88,32,6.4) (0.79,33,7.5) (1.06,21,3.3) (2.55,31,1.5)  \\
10 & 2.116-2.401 & 5 & 13.3$\sigma$ & (0.92,55,10.7) (1.01,40,6.8) (0.80,39,9.3) (1.11,26,3.8) (0.75,33,8.8)  \\
11 & 2.401-2.7535 & 5 & 14.1$\sigma$ & (0.90,49,8.4) (0.99,40,5.8) (0.78,42,9.1) (1.12,23,2.8) (0.85,24,4.6)  \\\hline
 & PC Mode & & & \\\hline
12 & 5.950-2.872 Msec & 4 & 7.1$\sigma$ & (0.80,109,0.003) (0.89,78,0.001) (0.69,96,0.003) (1.07,43,0.001) \\\hline
\end{tabular}
\end{center}
\label{tab:l060218s}
\end{table}

%% file: tab3.tex
\begin{table}[H]
\label{table:060218}
\begin{center}
\caption{Time-Integrated Spectroscopy for GRB~060218}
\vspace{5mm}
\footnotesize
\begin{tabular}{clcccc}\hline\hline
  Model & & $N_H$ & Norm.	& $\Gamma$ or kT (keV)  & $\chi^2/\nu$ \\\hline
	& & & & & \\
 WT Mode Data & $t=160$s-2.8ksec & & & & \\\hline
power-law       &        & 3.6 $\pm$ 0.1 & (3.43$\pm$ 0.01) $\times 10^{-9}$     & 1.84 $\pm$ 0.01 & 1740.56/768 \\
blackbody       &        & 0.4 $\pm$ 0.1 & (1.84$\pm$0.01 $\times 10^{-9}$    & 0.769 $\pm$ 0.002 & 14434.53/768 \\

power-law       &        & 7.2 $\pm$ 0.2 & (3.51$\pm$ 0.03 $\times 10^{-9}$     & 1.95  $\pm$ 0.01  & \\
  & $+$ blackbody        &               & (7.6$\pm$ 0.9) $\times 10^{-9}$     & 0.123 $\pm$ 0.002 &  1111.33/766 \\
power-law       &        & 4.0 $\pm$ 0.2 & (3.40$\pm$ 0.01) $\times 10^{-9}$   & 1.82  $\pm$ 0.01  & \\
  & $+$ MEKAL            &               & (8.0$\pm$ 0.5) $ \times 10^{-2}$    & 0.82 $\pm$ 0.02 & 1323.00/766 \\
	& & & & & \\
 PC Mode Data & $t=5.9$ksec-3.1Msec & & & & \\\hline
power-law      &         & 4.4 $\pm$ 0.3 & (3.1$\pm$0.3) $\times 10^{-13}$     & 3.4 $\pm$ 0.1 & 143.99/93 \\
blackbody      &         & 0.5 $\pm$ 0.2 & (9.0$\pm$0.4) $\times 10^{-14}$      & 0.36 $\pm$ 0.02 & 215.24/93 \\
power-law      &         & 6.9 $\pm$ 1.3 & (2.5$\pm$0.6) $\times 10^{-13}$     & 3.1 $\pm$ 0.2   & \\
     &    $+$ blackbody  &               & (2.9$\pm$0.7) $\times 10^{-12}$      & 0.10 $\pm$ 0.01 & 87.63/91 \\
power-law      &         & 5.6 $\pm$ 0.7 & (2.5$\pm$0.32) $\times 10^{-13}$    & 3.1 $\pm$ 0.1   & \\
     &    $+$ MEKAL      &               & (2.8$\pm$0.5)  $\times 10^{-4}$     & 0.23 $\pm$ 0.03 & 66.01/91\\\hline
\end{tabular}
\end{center}
{\footnotesize Notes: Column density $N_H$ measured in units of $10^{21}$ cm$^{-2}$.  $N_{H,\rm Galactic}=1.11 \times 10^{21}$ cm$^{-2}$ \citep{dickey1990}.  
Power-law model normalization (Norm.) in
units of erg cm$^{-2}$ s$^{-1}$ [0.5-10 keV].  Blackbody temperature measured in the source frame.
Blackbody normalization in units of erg cm$^{-2}$ s$^{-1}$.  MEKAL \citep{mekal} model normalization
in units of and equal to ${10^{-14} \over 4\pi D_A^2 (1+z)^2} \int n_e n_H dV$, for angular diameter distance $D_A$, electron density
$n_e$, and proton density $n_H$.}
\end{table}

%% file: tab4.tex
\begin{table}[H]
\begin{center}
\caption{Time-resolved Spectroscopy for GRB~050822}
\vspace{5mm}
\tiny
\begin{tabular}{cccccccccc}\hline\hline
 \# &  Time (s) & BB Flux & $kT$ (keV) & $N_H$  & $\chi^2/\nu$ & PL Flux & $\Gamma$ & $N_H$ & $\chi^2/\nu$ \\\hline
 & WT Mode & & & & & & & & \\\hline
 1 & 111.0-121.0 & 26.1 $\pm$ 1.5 & 0.66 $\pm$ 0.03 & 0.1 $\pm$ 0.1 & 105.93/59 & 44.6 $\pm$ 3.4 & 1.8 $\pm$ 0.2 & 1.9 $\pm$ 0.6 & 48.50/59 \\
 2 & 121.0-131.0 & 28.0 $\pm$ 1.6 & 0.63 $\pm$ 0.03 & 0.1 $\pm$ 0.1 & 107.13/66 & 51.1 $\pm$ 4.0 & 2.0 $\pm$ 0.2 & 2.7 $\pm$ 0.6 & 53.64/66 \\
 3 & 131.0-141.0 & 26.1 $\pm$ 1.4 & 0.57 $\pm$ 0.03 & 0.1 $\pm$ 0.1 & 134.80/65 & 50.0 $\pm$ 4.2 & 2.1 $\pm$ 0.2 & 2.5 $\pm$ 0.6 & 64.32/65 \\
 4 & 141.0-151.0 & 20.5 $\pm$ 1.3 & 0.54 $\pm$ 0.03 & 0.1 $\pm$ 0.1 & 103.56/54 & 38.6 $\pm$ 3.2 & 2.0 $\pm$ 0.2 & 1.9 $\pm$ 0.6 & 32.66/54 \\
 5 & 151.0-160.9 & 16.5 $\pm$ 1.2 & 0.47 $\pm$ 0.03 & 0.1 $\pm$ 0.1 & 85.38/46 & 33.6 $\pm$ 3.9 & 2.4 $\pm$ 0.3 & 2.4 $\pm$ 0.7 & 33.79/46 \\
 6 & 160.9-170.9 & 10.1 $\pm$ 0.8 & 0.40 $\pm$ 0.04 & 0.1 $\pm$ 0.1 & 80.37/31 & 18.2 $\pm$ 2.6 & 2.2 $\pm$ 0.4 & 1.2 $\pm$ 0.9 & 44.87/31 \\
 7 & 170.9-200.9 & 3.8 $\pm$ 0.3 & 0.35 $\pm$ 0.03 & 0.1 $\pm$ 0.1 & 91.20/36 & 6.4 $\pm$ 0.8 & 2.3 $\pm$ 0.3 & 0.7 $\pm$ 0.7 & 33.05/36 \\
 8 & 200.9-240.8 & 3.6 $\pm$ 0.3 & 0.37 $\pm$ 0.02 & 0.1 $\pm$ 0.1 & 80.67/44 & 7.3 $\pm$ 1.2 & 2.7 $\pm$ 0.3 & 2.2 $\pm$ 0.7 & 39.59/44 \\
 9 & 240.8-280.8 & 2.7 $\pm$ 0.2 & 0.35 $\pm$ 0.03 & 0.1 $\pm$ 0.1 & 47.69/32 & 5.5 $\pm$ 1.3 & 3.1 $\pm$ 0.4 & 2.2 $\pm$ 0.8 & 24.38/32 \\
 10 & 280.8-429.6 & 2.1 $\pm$ 0.2 & 0.34 $\pm$ 0.02 & 0.1 $\pm$ 0.1 & 76.08/52 & 4.6 $\pm$ 1.0 & 3.1 $\pm$ 0.3 & 2.4 $\pm$ 0.7 & 51.37/52 \\
 11 & 429.6-449.6 & 10.3 $\pm$ 2.3 & 0.26 $\pm$ 0.02 & 0.7 $\pm$ 0.6 & 33.37/43 & 81.1 $\pm$ 44.3 & 5.2 $\pm$ 0.5 & 5.6 $\pm$ 1.0 & 30.33/43 \\
 12 & 449.6-469.5 & 10.0 $\pm$ 2.0 & 0.25 $\pm$ 0.02 & 0.4 $\pm$ 0.5 & 35.46/46 & 99.6 $\pm$ 54.3 & 5.6 $\pm$ 0.5 & 5.7 $\pm$ 1.0 & 41.03/46 \\
 13 & 469.5-489.5 & 10.2 $\pm$ 4.1 & 0.20 $\pm$ 0.02 & 0.8 $\pm$ 0.8 & 35.17/37 & 169.7 $\pm$ 72.5 & 6.7 $\pm$ 0.8 & 6.6 $\pm$ 1.4 & 35.03/37 \\
 14 & 489.5-509.4 & 6.4 $\pm$ 0.7 & 0.21 $\pm$ 0.02 & 0.1 $\pm$ 0.4 & 51.06/33 & 53.3 $\pm$ 11.1 & 5.9 $\pm$ 0.9 & 4.8 $\pm$ 1.6 & 54.71/33 \\
 15 & 509.4-549.3 & 4.1 $\pm$ 0.4 & 0.17 $\pm$ 0.02 & 0.3 $\pm$ 0.6 & 31.63/31 & 23.6 $\pm$ 6.6 & 6.1 $\pm$ 1.0 & 4.2 $\pm$ 1.5 & 37.11/31 \\\hline
 & PC Mode & & & & & & & & \\\hline
 16 & 354.4-12823 & 0.45 $\pm$ 0.03 & 0.27 $\pm$ 0.02 & 0.1 $\pm$ 0.1 & 145.96/38 & 0.81 $\pm$ 0.09 & 2.1 $\pm$ 0.2 & 0.6 $\pm$ 0.2 & 52.01/38 \\
 17 & 12823-34706 & 0.19 $\pm$ 0.02 & 0.39 $\pm$ 0.05 & 0.1 $\pm$ 0.1 & 107.85/34 & 0.40 $\pm$ 0.04 & 2.1 $\pm$ 0.3 & 1.3 $\pm$ 0.6 & 45.42/34 \\
 18 & 34706-88110 & 0.04 $\pm$ 0.01 & 0.43 $\pm$ 0.04 & 0.1 $\pm$ 0.1 & 72.60/32 & 0.08 $\pm$ 0.01 & 2.2 $\pm$ 0.3 & 1.6 $\pm$ 0.6 & 31.15/32 \\\hline
 & PC $+$ WT & & & & & & & & \\\hline
 19 & 549.3-1000.0 & 1.2 $\pm$ 0.3 & 0.17 $\pm$ 0.02 & 0.3 $\pm$ 0.1 & 48.84/19 & 12.6 $\pm$ 1.3 & 4.0 $\pm$ 1.1 & 1.2 $\pm$ 1.4 & 35.12/19 \\\hline
\end{tabular}
\end{center}
{\footnotesize Notes: Column density $N_H$ measured in units of $10^{21}$ cm$^{-2}$.  $N_{H,\rm Galactic}=2.25 \times 10^{20}$ cm$^{-2}$ \citep{dickey1990}.
Power-law (PL) and blackbody (BB) fluxes measured in $10^{-10}$ erg cm$^{-2}$ s$^{-1}$, 0.5-10 keV band for PL model. Blackbody temperature measured
in the observer frame.}
\label{tab:050822}
\end{table}

%% file: tab5.tex
\begin{table}[H]
\begin{center}
\caption{Time-resolved Line Search for GRB~050822}
\vspace{5mm}
\footnotesize
\begin{tabular}{ccccc}\hline\hline
 \# &  Time  & N  & Signif. & Line Energy , Equivalent Width , Flux \\
    &   [s]   &  lines            &     &  (keV,eV,$10^{-11}$ erg cm$^{-2}$ s$^{-1}$) \\\hline
 & WT Mode & & & \\\hline
 1 & 111.0-121.0 & 0 & $<1.8\sigma$ & ... \\
 2 & 121.0-131.0 & 0 & $<1.7\sigma$ & ... \\
 3 & 131.0-141.0 & 3 & $2.1\sigma$ & (1.94,93,7.7) (2.91,115,5.9) (3.66,192,7.6) \\
 4 & 141.0-151.0 & 0 & $<1.2\sigma$ & ... \\
 5 & 151.0-160.9 & 2 & $2.0\sigma$ & (1.01,49,5.9) (1.11,51,5.5) \\
 6 & 160.9-170.9 & 4 & $2.7\sigma$ & (0.66,90,7.6) (0.88,67,4.2) (1.13,111,5.5) (1.29,95,4.1) \\
 7 & 170.9-200.9 & 0 & $<1.9\sigma$  & ... \\
 8 & 200.9-240.8 & 0 & $<1.5\sigma$ & ... \\
 9 & 240.8-280.8 & 0 & $<0.6\sigma$ & ... \\
 10 & 280.8-429.6 & 0 & $<1.2\sigma$ & ... \\
 11 & 429.6-449.6 & 0 & $<0.6\sigma$ & ... \\
 12 & 449.6-469.5 & 0 & $<1.3\sigma$ & ... \\
 13 & 469.5-489.5 & 0 & $<1.6\sigma$ & ... \\
 14 & 489.5-509.4 & 5 & $4.4\sigma$ & (0.81,82,3.6) (0.91,142,4.5) (1.04,194,4.3) (1.23,221,3.1) (1.49,265,2.2) \\
 15 & 509.4-549.3 & 0 & $<1.9\sigma$ & ... \\\hline
 & PC Mode & & & \\\hline
 16 & 354.4-12823 & 3 & $2.5\sigma$ & (0.41,75,0.6) (0.63,106,0.5) (3.48,457,0.3) \\
 17 & 12823-34706 & 0 & $<1.9\sigma$ & ... \\
 18 & 34706-88110 & 0 & $<2.0\sigma$ & ... \\\hline
 & PC $+$ WT & & & \\\hline
 19 & 549.3-1000.0 & 5 & $3.0\sigma$ & (0.65,131,1.1) (0.75,64,0.4) (0.82,65,0.3) (0.90,65,0.3) (1.09,50,0.2) \\\hline
\end{tabular}
\end{center}
\label{tab:l050822}
\end{table}

%% file: tab6.tex
\begin{table}[H]
\begin{center}
\caption{Time-resolved Spectroscopy for GRB~050714B}
\vspace{5mm}
\tiny
\begin{tabular}{cccccccccc}\hline\hline
 \# &  Time Reg. (s) & BB Flux & $kT$ (keV) & $N_H$  & $\chi^2/\nu$ & PL Flux & $\Gamma$ & $N_H$ & $\chi^2/\nu$ \\\hline
  & WT Mode & & & & & & & & \\\hline
1 & 157.4-189.7   & 7.1 $\pm$ 1.5 & 0.23 $\pm$ 0.02 & 2.66 $\pm$ 0.59 & 33.49/31 & 85.49 $\pm$ 40.36 & 5.90 $\pm$ 0.44 & 0.81 $\pm$ 0.09 & 29.39/31 \\\hline
  & PC Mode & & & & & & & & \\\hline
2 & 221.4-58.3ks  & 0.006 $\pm$ 0.001 & 0.30 $\pm$ 0.03 & 0.30 $\pm$ 0.23 & 97.19/35 & 0.015 $\pm$ 0.003 & 3.24 $\pm$ 0.28 & 0.31 $\pm$ 0.05 & 76.78/35 \\
  &               & 0.033 $\pm$ 0.007 & 0.13 $\pm$ 0.03 & 3.8 $\pm$ 1.4 &   $+$     & 0.008 $\pm$ 0.002 & 2.2 $\pm$ 0.5  & tied           & 48.64/33 \\
3 & 275.0-525.0   & 0.74 $\pm$ 0.37 & 0.22 $\pm$ 0.03 & 0.83 $\pm$ 0.69 & 35.67/19 & 4.28 $\pm$ 2.69 & 5.11 $\pm$ 0.63 & 0.53 $\pm$ 0.10 & 39.92/19 \\
  &               & 7.8 $\pm$ 0.4 & 0.11 $\pm$ 0.03 & 4.5 $\pm$ 2.3 &   $+$     & $0.7 \pm 0.1$ & 3.1 $\pm$ 1.1  & tied           & 21.40/17 \\\hline
\end{tabular}
\end{center}
{\footnotesize Notes: Column density $N_H$ measured in units of $10^{21}$ cm$^{-2}$.  $N_{H,\rm Galactic}=5.31 \times 10^{20}$ cm$^{-2}$ \citep{dickey1990}.  
Flux measured in units of $10^{-10}$ erg cm$^{-2}$ s$^{-1}$, 0.5-10 keV
for the power-law (PL) and bolometric for the blackbody (BB).  Blackbody temperature measured in the observer frame.}
\label{tab:050714B}
\end{table}

%% file: tab7.tex
\begin{table}[H]
\begin{center}
\caption{Time-resolved Line Search for GRB~050714B}
\vspace{5mm}
\footnotesize
\begin{tabular}{ccccc}\hline\hline
 \# &  Time  & N  & Signif. & Line Energy , Equivalent Width , Flux \\
    &  [s]     &  lines      &     &  (keV,eV,$10^{-12}$ erg cm$^{-2}$ s$^{-1}$) \\\hline
 & WT Mode & & & \\\hline
 1 & 157.4-189.7 & 0 & $<$1.8$\sigma$ & ... \\\hline
 & PC Mode & & & \\\hline
 2 & 221.4-58.3 ks & 4 & 4.2$\sigma$ & (0.91,147,0.09) (0.76,141,0.12) (0.56,238,0.35) (1.12,64,0.03)  \\
 3 & 275.0-525.0 & 4 & 3.4$\sigma$ & (0.74,170,57.5) (0.59,340,261) (0.93,108,16.0) (0.83,68,15.4) \\\hline
\end{tabular}
\end{center}
\label{tab:l050714B}
\end{table}

%% file: tab8.tex
\begin{table}[H]
\begin{center}
\caption{Spectroscopy of Bright ($>10$ cts/s peak) XRT Flares}
\vspace{5mm}
\footnotesize
\begin{tabular}{cccccc}\hline\hline
  Burst & Time Reg. (s) & Fluence & $\Gamma$ & $N_H$ & $\chi^2/\nu$ \\
         &               & ($10^{-7}$ erg cm$^{-2}$) & & ($10^{21}$ cm$^{-2}$) & \\\hline
050502b & 400-1200 & 6.61 $\pm$ 0.09 & 2.43 $\pm$ 0.03 & 1.76 $\pm$ 0.07 & 373.28/357 \\
050607 & 250-600 & 0.38 $\pm$ 0.03 & 2.29 $\pm$ 0.13 & 2.11 $\pm$ 0.36 & 30.78/34 \\
050712 & 150-300 & 0.55 $\pm$ 0.02 & 2.06 $\pm$ 0.10 & 2.35 $\pm$ 0.32 & 91.62/73 \\
050713A & 95-150 & 3.07 $\pm$ 0.10 & 2.31 $\pm$ 0.06 & 5.55 $\pm$ 0.25 & 216.45/207 \\
050714B & 275-525 & 1.07 $\pm$ 0.67 & 5.11 $\pm$ 0.63 & 5.25 $\pm$ 1.02 & 39.92/19 \\
050730 & 130-300 & 1.43 $\pm$ 0.04 & 1.63 $\pm$ 0.05 & 1.69 $\pm$ 0.21 & 178.00/171 \\
050730 & 300-600 & 1.90 $\pm$ 0.04 & 1.72 $\pm$ 0.05 & 1.02 $\pm$ 0.14 & 215.10/227 \\
050730 & 600-800 & 0.84 $\pm$ 0.03 & 1.84 $\pm$ 0.07 & 0.43 $\pm$ 0.18 & 127.01/134 \\
050820 & 215-252 & 2.40 $\pm$ 0.06 & 0.99 $\pm$ 0.05 & 1.87 $\pm$ 0.23 & 202.11/207 \\
050822 & 410-650 & 4.03 $\pm$ 0.50 & 4.95 $\pm$ 0.13 & 4.07 $\pm$ 0.24 & 141.36/127 \\
050904 & 350-600 & 1.37 $\pm$ 0.03 & 1.73 $\pm$ 0.05 & 1.49 $\pm$ 0.15 & 221.51/227 \\
051117A & 1250-1725 & 6.81 $\pm$ 0.09 & 2.31 $\pm$ 0.03 & 1.94 $\pm$ 0.08 & 368.05/364 \\
051117A & 800-1250 & 4.77 $\pm$ 0.08 & 2.32 $\pm$ 0.03 & 1.76 $\pm$ 0.09 & 333.41/307 \\
060111A & 200-500 & 5.59 $\pm$ 0.09 & 2.29 $\pm$ 0.03 & 2.70 $\pm$ 0.10 & 355.60/335 \\
060124 & 300-650 & 16.48 $\pm$ 0.13 & 1.39 $\pm$ 0.02 & 2.36 $\pm$ 0.07 & 611.60/611 \\
060124 & 650-900 & 12.91 $\pm$ 0.13 & 1.92 $\pm$ 0.02 & 2.26 $\pm$ 0.08 & 486.54/453 \\
060204B & 100-270 & 1.95 $\pm$ 0.05 & 2.00 $\pm$ 0.05 & 2.78 $\pm$ 0.19 & 168.65/201 \\
060204B & 270-450 & 0.25 $\pm$ 0.02 & 2.68 $\pm$ 0.13 & 2.88 $\pm$ 0.34 & 78.40/70 \\
060210 & 100-165 & 1.32 $\pm$ 0.03 & 1.82 $\pm$ 0.06 & 2.04 $\pm$ 0.20 & 149.31/157 \\
060210 & 165-300 & 2.70 $\pm$ 0.05 & 1.92 $\pm$ 0.04 & 1.98 $\pm$ 0.14 & 210.08/260 \\
060210 & 350-450 & 1.18 $\pm$ 0.07 & 2.93 $\pm$ 0.08 & 2.86 $\pm$ 0.21 & 127.92/141 \\
060312 & 100-200 & 0.14 $\pm$ 0.01 & 2.30 $\pm$ 0.09 & 2.99 $\pm$ 0.29 & 72.11/105 \\\hline
\end{tabular}
\end{center}
\label{tab:flares}
\end{table}

%% file: tab9.tex
\begin{table}[H]
\begin{center}
\caption{Line Detections in the Bright ($>10$ cts/s peak) XRT Flares}
\vspace{5mm}
\footnotesize
\begin{tabular}{cccccccccc}\hline\hline
 Burst &  Time & N  & Signif. & Line Energy , Equivalent Width , Flux \\
     &    [s]  &  lines          &     &  (keV,eV,$10^{-11}$ erg cm$^{-2}$ s$^{-1}$) \\\hline
050502b & 400-1200 & 4 & 3.6$\sigma$ & (1.44,15,0.4) (1.69,18,0.4) (4.48,57,0.3) (2.14,17,0.2) \\
050607  & 250-600  & 5 & 2.1$\sigma$ & (2.11,152,0.2) (0.79,62,0.4) (0.61,118,1.1) (3.29,409,0.3) (2.57,216,0.3) \\
050712  & 150-300  & 0 & $<$1.9$\sigma$ & ...\\
050713A & 95-150   & 3 & 2.9$\sigma$ & (1.82,40,4.1) (4.48,97,3.0) (3.72,65,2.6) \\
050714B & 275-525  & 3 & 3.4$\sigma$ & (0.74,162,6.9) (0.59,301,30.7) (0.93,98,1.7) \\
050730  & 130-300  & 1 & 2.2$\sigma$ & (1.97,53,0.7) \\
050730  & 300-600  & 5 & 2.9$\sigma$ & (1.12,31,0.4) (1.41,29,0.3) (1.01,27,0.4) (1.87,47,0.5) (0.77,18,0.3) \\
050730  & 600-800  & 0 & $<$1.9$\sigma$ & ...\\
050820  & 215-252  & 2 & 2.3$\sigma$ & (4.27,72,4.9) (1.96,41,2.8) \\
050822  & 410-650  & 3 & 3.9$\sigma$ & (0.91,22,2.8) (0.61,44,30.9) (2.06,76,0.3) \\
050904  & 350-600  & 0 & $<$1.7$\sigma$ & ...\\
051117A & 1250-1725 & 3 & 4.4$\sigma$ & (1.88,34,0.9) (2.83,26,0.4) (0.60,25,3.0) \\
051117A & 800-1250 & 4 & 4.3$\sigma$ & (1.75,30,0.6) (1.37,20,0.5) (0.92,16,0.7) (1.05,12,0.4) \\
060111A & 200-500  & 4 & 2.3$\sigma$ & (1.80,20,0.7) (1.04,12,0.9) (0.90,11,0.9) (0.80,10,1.0) \\
060124  & 300-650  & 3 & 3.2$\sigma$ & (1.89,18,1.2) (4.57,29,1.4) (3.47,18,0.9) \\
060124  & 650-900  & 5 & 2.6$\sigma$ & (3.20,28,1.5) (4.51,40,1.6) (4.89,42,1.6) (1.12,8,1.2) (2.73,18,1.1) \\
060204B & 100-270  & 4 & 2.8$\sigma$ & (2.64,61,1.7) (2.08,44,1.5) (2.83,56,1.4) (4.32,83,1.4) \\
060204B & 270-450  & 0 & $<$1.8$\sigma$ & ...\\
060210  & 100-165  & 0 & $<$1.6$\sigma$ & ...\\
060210  & 165-300  & 0 & $<$1.3$\sigma$ & ...\\
060210  & 350-450  & 1 & 2.1$\sigma$ & (0.62,50,8.7) \\
060312  & 100-200  & 1 & 2.5$\sigma$ & (0.60,117,1.7) \\\hline
\end{tabular}
\end{center}
\label{tab:flaresl}
\end{table}

%% file: tab10.tex
\begin{table}[H]
\begin{center}
\caption{Afterglow Energetics Parameters}
\vspace{5mm}
\footnotesize
\begin{tabular}{cccccccccc}\hline\hline
  Burst & $z$   &       $E_{\gamma,{\rm iso}}$ &        $L_{{\rm X},t=10{\rm hrs}}$  &     $S_{\rm BB}$            & $\delta t_{\rm BB} [s]$ \\
        &       &       [$10^{51}$ erg]       &    [$10^{44}$ erg s$^{-1}$]  &    [$10^{51}$ erg] & [s] \\\hline
060218  & 0.033 &       $0.062\pm0.003$       & $0.013\pm0.003$              & $0.23\pm0.07$               & $\sim 300$ \\
060202  & 0.783 &       $4.4\pm0.4$           & $43\pm11$                  & $1.3\pm0.4$               & $\lessim 1000$ \\
050822  & 1.2   &       $15\pm1$              & $410\pm80$                   & $0.43\pm0.18$               & $\sim 100$ \\
050714B & 2.66  &       $13\pm2$              & $440\pm90$                   & $2.8\pm1.0$                 & $\sim 200$ \\\hline
\end{tabular}
\end{center}
{\footnotesize Notes: Redshifts $z$ for GRBs 060202, 050822, and 050714B are inferred from the X-ray spectroscopy (Section \ref{sec:bigthree}).}
\label{tab:aft}
\end{table}